
\input harvmac
\overfullrule=0pt
\def\Title#1#2{\rightline{#1}\ifx\answ\bigans\nopagenumbers\pageno0\vskip1in
\else\pageno1\vskip.8in\fi \centerline{\titlefont #2}\vskip .5in}

%
%
\baselineskip=18pt plus 2pt minus 2pt
\def\sq{{\vbox {\hrule height 0.6pt\hbox{\vrule width 0.6pt\hskip 3pt
   \vbox{\vskip 6pt}\hskip 3pt \vrule width 0.6pt}\hrule height 0.6pt}}}
\def\CD{{\cal D}}
\font\ticp=cmcsc10
\def\sq{{\vbox {\hrule height 0.6pt\hbox{\vrule width 0.6pt\hskip 3pt
   \vbox{\vskip 6pt}\hskip 3pt \vrule width 0.6pt}\hrule height 0.6pt}}}
\def\ajou#1&#2(#3){\ \sl#1\bf#2\rm(19#3)}
\def\frac#1#2{{#1 \over #2}}

\def\pmi{{\partial_-}}
\def\p+{{\partial_+}}

\def\air{{\rightarrow}}
\def\ci{{\cal I}}
\def\half{{1 \over 2}}
\ifx\epsfbox\UnDeFiNeD\message{(NO epsf.tex, FIGURES WILL BE IGNORED)}
\def\figin#1{\vskip2in}
\else\message{(FIGURES WILL BE INCLUDED)}\def\figin#1{#1}\fi
\def\ifig#1#2#3{\xdef#1{fig.~\the\figno}
\goodbreak\midinsert\figin{\centerline{#3}}%
\smallskip\centerline{\vbox{\baselineskip12pt
\advance\hsize by -1truein\noindent\footnotefont{\bf Fig.~\the\figno:} #2}}
\bigskip\endinsert\global\advance\figno by1}
%
%
\lref\cham{A. Chamseddine, \ajou  Nucl. Phys. &B368 (92) 98.}
\lref\andywaffle{A. Strominger, unpublished.}
\lref\bizon{P. Bizon and R. Wald, \ajou Phys. Lett. &B267 (91) 173.}
\lref\qtdg{S. Giddings and A. Strominger, ``Quantum theories of dilaton
gravity,'' preprint UCSBTH-92-28, hep-th/9207034.}
\lref\chargetwod{O. Lechtenfeld and C. Nappi, Institute for Advanced
Study preprint IASSNS-HEP-92-22,
hep-th/9204026 \semi S. Nojiri and I. Oda, preprints NDA-FP-5-92,
hep-th/9207077 and NDA-FP-4-92, hep-th/9206087.}
\lref\dgt{F. Dowker, R. Gregory and J. Traschen, ``Euclidean black hole
vortices,'' \ajou Phys. Rev. &D45 (92) 2762.}
\lref\hairrev{S. Coleman, J. Preskill and F. Wilczek, ``Quantum hair
on black holes,'' \ajou Nucl. Phys. &B378
(92) 175.}
\lref\luckock{H. Luckock and I. Moss, \ajou Phys. Lett. &176B (86) 341.}
\lref\bd{N. D. Birrell and P. C. W. Davies, {\it Quantum fields in
curved space},
Cambridge University Press, Cambridge 1982.}
\lref\suss{A. Peet, L. Susskind and L. Thorlacius, Stanford
preprint SU-ITP-92-16 (1992).}
\lref\bek{J. D. Bekenstein, ``Black holes and entropy,'' \ajou Phys. Rev. &D7
 (73) 2333 \semi ``Generalized second law of thermodynamics in black-hole
 physics,'' \ajou Phys. Rev. &D9 (74) 3292.}
\lref\host{G. T. Horowitz and A. Strominger, ``Black Strings and p-Branes,''
\ajou Nucl. Phys. &B360 (91) 197.}
\lref\chs{C. Callan, J. Harvey and A. Strominger, ``Supersymmetric String
Solitons,'' in {\it String theory and quantum gravity '91}, Proceedings of the
Trieste Spring School 1991, eds. J. Harvey,
R. Iengo, K. Narain, S. Randjbar-Daemi and H. Verlinde, World Scientific
(1991).}
\lref\gist{S. Giddings and A, Strominger, ``Exact Black Fivebranes,''
\ajou Phys. Rev. Lett. &67 (91) 2930.}
\lref\psstw{J. Preskill, P. Schwarz, A. Shapere, S. Trivedi and F. Wilczek,
``Limitations on the statistical description of black holes,'' \ajou
Mod. Phys. Lett. &A6 (91) 2353.}
\lref\ddy{A. Shapere, S. Trivedi and F. Wilczek, \ajou Mod. Phys. Lett. &A6
(91) 2677. }
\lref\poly{A. M. Polyakov, ``Quantum geometry of bosonic strings,''
\ajou Phys. Lett. &103B (81) 207.}
\lref\dmass{R. Gregory and J. Harvey, to appear and
J. Horne and G. Horowitz, to appear.}
\lref\cash{Y. Aharonov, A. Casher and S. Nussinov, \ajou Phys. Lett. &191B
(87) 51.}
\lref\walddewit{R. M. Wald, ``Black holes, singularities,
and predictability,'' in Quantum  Theory of Gravity, S. M. Christensen ed.
(Adam Hilger, Bristol  U.K. 1984). }
\lref\corn{T. Banks and M. O'Loughlin, ``Classical and quantum production
of cornucopians at energies below $10^{18}$ GeV'', Rutgers
preprint RU-92-14 (1992).}
\lref\bartnik{R. Bartnik and J. McKinnon, \ajou Phys. Rev. Lett. &61 (88)
 141 \semi P. Bizon, \ajou Phys. Rev. Lett. &64 (90) 2844 \semi
 M. Volkov and D. Gal'tsov, \ajou Sov. J. Nucl. Phys. &51 (90) 1171 \semi
 H. Kunzle and A. Masood-ul-Alam, \ajou J. Math. Phys. &31 (90) 928. }
\lref\gihu{G. Gibbons and C. M. Hull, \ajou Phys. Lett. &109B (82) 190.}
\lref\bunwa{R. M. Wald, 1991 Erice Lectures (unpublished). }
\lref\hortr{G. Horowitz, in {\it String theory and quantum gravity '91},
Proceedings of the Trieste Spring School 1991,
 eds. J. Harvey, R. Iengo, K. Narain, S. Randjbar-Daemi and
 H. Verlinde, World Scientific (1991).}
\lref\Mann{R. B. Mann, ``Lower dimensional gravity,'' in Proc. of 4th Canadian
Conf. on General Relativity and Relativistic Astrophysics, Winnipeg Canada
and references therein.}
\lref\renata{R. Kallosh, A. Linde, T. Ort\'in, A. Peet and A. Van Proeyen,
``Supersymmetry as a cosmic censor,'' preprint SU-ITP-92-13 (1992).}
\lref\gary{J. Horne and G. Horowitz, ``Rotating dilaton black holes,'' preprint
UCSBTH-92-11, hep-th/9203083 (1992) \semi A. Sen, ``Rotating charged
black hole solution in heterotic string theory,'' Tata preprint
TIFR/TH/92-20.}
\lref\gm{G. Gibbons and K. Maeda, \ajou Nucl. Phys. &B298 (88) 741.}
\lref\ghs{D. Garfinkle, G. Horowitz, and A. Strominger, \ajou Phys. Rev. &D43
(91) 3140; Erratum: \ajou Phys. Rev. &D45 (92) 3888.}
\lref\toupee{L. Krauss and F. Wilczek, \ajou Phys. Rev. Lett. &62 (89)
 1221 \semi S. Coleman, J. Preskill and F. Wilczek, \ajou Phys.
 Rev. Lett. &67 (91) 1975 \semi S. Coleman, J. Preskill and F. Wilczek,
``Quantum hair on black holes,'' \ajou Nucl. Phys. &B378 (92) 175 \semi
 F. Dowker, R. Gregory and J. Traschen, ``Euclidean black hole
vortices,'' \ajou Phys. Rev. &D45 (92) 2762.}
\lref\hair{M. Bowick, S. Giddings, J. Harvey, G. Horowitz, and A.
Strominger, \ajou Phys. Rev. Lett. &61 (88) 2823.}
\lref\kw{L. Krauss and F. Wilczek, \ajou Phys. Rev. Lett. &62 (89) 1221.}
\lref\monhole{K. Lee, V. P. Nair and E. J. Weinberg, \ajou Phys. Rev. &D45
(92) 2751.}
\lref\wald{R. M. Wald, {\it General Relativity},
The University of Chicago Press,
Chicago (1984). }
\lref\mik{A. Mikovic, ``Exactly solvable
models of 2d dilaton quantum gravity,''Queen Mary preprint QMW/PH/92/12,
hep-th/9207006.}
\lref\ham{ K. Hamada, ``Quantum theory of dilaton gravity in 1+1 dimensions,''
preprint UT-Komaba 92-7, hep-th/9206071}
\lref\tbo{T. Banks and M. O'Loughlin, ``Two-dimensional quantum gravity
 in Minkowski space,'' \ajou Nucl. Phys. &B362 (91) 649.}
\lref\jptb{S. R. Das, S. Naik and S. R. Wadia, ``Quantization of the
Liouville mode and string theory,'' \ajou Mod. Phys. Lett. &A4 (89) 1033 \semi
J. Polchinski, ``A two-dimensional model for quantum
gravity,'' \ajou Nucl. Phys. &B324 (89) 123 \semi
T. Banks and J. Lykken, ``String theory and two-dimensional quantum
gravity,'' \ajou Nucl. Phys. &B331 (90) 173.}
\lref\rut {J. G. Russo and A. A. Tseytlin,  ``Scalar-Tensor Quantum Gravity
in Two Dimensions,''  preprint SU-ITP-92-2, DAMTP-1-1992.}
\lref\hver{H. Verlinde,
``Black
holes and strings in two dimensions,''
in the proceedings of the Sixth Marcel Grossman Meeting,
World Scientific, Singapore (1992).}
\lref\RuTs{J.G. Russo and A.A. Tseytlin, ``Scalar-tensor quantum gravity
in two dimensions,'' Stanford/Cambridge preprint SU-ITP-92-2=DAMTP-1-1992.}
\lref\Mandal{G. Mandal, A Sengupta, and S. Wadia, ``Classical solutions of
two-dimensional
string theory,'' \ajou Mod. Phys. Lett. &A6 (91) 1685.}
\lref\WittTwod{E. Witten, ``On string theory and black holes,''\ajou Phys. Rev&
D44 (91) 314.}
\lref\CGHS{C.G. Callan, S.B. Giddings, J.A. Harvey, and A. Strominger,
``Evanescent black holes,"\ajou Phys. Rev. &D45 (92) R1005.}
\lref\BDDO{T. Banks, A. Dabholkar, M.R. Douglas, and M O'Loughlin, ``Are
horned particles the climax of Hawking evaporation?'' \ajou Phys. Rev.
&D45 (92) 3607.}
\lref\rst{J.G. Russo, L. Susskind, and L. Thorlacius, ``The
Endpoint of Hawking Evaporation,'' Stanford preprint SU-ITP-92-17.}
\lref\RST{J.G. Russo, L. Susskind, and L. Thorlacius, ``Black hole
evaporation in 1+1 dimensions,'' Stanford preprint SU-ITP-92-4.}
\lref\Stro{A. Strominger, ``Fadeev-Popov ghosts and 1+1 dimensional black
hole evaporation,'' {\it Phys. Rev. D}, to appear, hep-th/9205028.}
\lref\deAli{S.P. deAlwis, ``Quantization of a theory of 2d dilaton
gravity,'' Boulder preprint COLO-HEP-280, hep-th/9205069,
 ``Black hole physics from Liouville theory,''
Boulder preprint COLO-HEP-284, hep-th/9206020,'' Quantum Black Holes in Two
Dimensions,'' Boulder preprint COLO-HEP-288, hep-th/9207095.}
\lref\BiCa{A. Bilal and C. Callan, ``Liouville models of black hole
evaporation,'' Princeton preprint PUPT-1320, hep-th/9205089.}
\lref\GiStunpub{S.B. Giddings and A. Strominger, unpublished.}
\lref\CrFu{S. M. Christensen and S. A. Fulling, ``Trace anomalies and the
Hawking effect,''\ajou Phys. Rev. &D15 (77) 2088.}
\lref\GiNe{S.B. Giddings and W.M. Nelson, ``Quantum emission from
two-dimensional black holes,'' UCSB preprint UCSBTH-92-15,
hep-th/9204072.}
\lref\Hawk{S.W. Hawking, ``Evaporation of two dimensional black holes,''
\ajou Phys. Rev. Lett. &69 (92) 406.}
\lref\SuTh{L. Susskind and L. Thorlacious, ``Hawking radiation and
back-reaction,'' Stanford preprint SU-ITP-92-12, hep-th/9203054.}
\lref\BGHS{B. Birnir, S.B. Giddings, J.A. Harvey, and A. Strominger,
``Quantum black holes,'' \ajou Phys. Rev.  &D46 (92) 638. }
\lref\rom{R. Jackiw, ``Liouville field theory: a two-dimensional model for
gravity?,'' in {\sl Quantum theory of gravity}, S. Christensen, ed.
(Hilger, Bristol U.K. 1984); D. Cangemi and R. Jackiw ``Gauge Invariant
Formulations of Lineal Gravity'' MIT preprint CTP-2085 (1992).}
\lref\tei{C. Teitelboim `` The hamiltonian structure of spacetime and its
relation with the conformal anomaly,'' in {\sl Quantum theory of gravity}, S.
Christensen, ed.
(Hilger, Bristol U.K. 1984)}
\lref\pnrs{See e.g. S. W. Hawking and G. F. R. Ellis, {\it The large scale
structure of space-time}, Cambridge University Press (1973).}
\lref\Hawktwo{S. W. Hawking, \ajou Phys. Rev. &D14 (76) 2460.}
\lref\steve{S. Giddings, ``Black Holes and Massive Remnants,'' preprint
UCSBTH-92-09, hep-th/9203059.}
\lref\page{D. Page, \ajou Phys. Rev. Lett. &44 (80) 301.}
\lref\dxbh{S. B. Giddings and A. Strominger, ``Dynamics
of Extremal Black Holes'',\ajou Phys. Rev. &D46 (92) 627.}
\lref\tHooft{G. 't Hooft, \ajou Nucl. Phys. &B335 (90) 138.}
\lref\hwhat{S. W. Hawking and J. M. Stewart, ``Naked and thunderbolt
singularities in black hole evaporation,'' hep-th/9207105.}
\lref\asup{A. Strominger, unpublished.}
\lref\malf{M. Alford and A. Strominger, ``S-Wave Scattering of Charged Fermions
by a Magnetic Black Hole,'' \ajou Phys. Rev. Lett. &69 (92) 563.}
\lref\Hawkfirst{S. W. Hawking,\ajou Commun. Math. Phys. &43 (75) 199 \semi
\ajou Phys. Rev. &D14 (76) 2460.}
\lref\past{Y. Park and A. Strominger, to appear.}

\Title{\vbox{\baselineskip12pt\hbox{EFI-92-41}
\hbox{hep-th/9209055}
}}
{\vbox{\centerline {Quantum Aspects of Black Holes}
}}

\centerline{{\ticp Jeffrey A. Harvey}\footnote{$^\dagger$}
{Email address:
harvey@yukawa.uchicago.edu}}
\vskip.1in
\centerline{\sl Enrico Fermi Institute}
\centerline{\sl University of Chicago}
\centerline{\sl 5640 Ellis Avenue, Chicago, IL 60637}
\vskip .1in
\centerline{{\ticp Andrew Strominger}\footnote{$^*$}
{Email addresses:
andy@denali.physics.ucsb.edu, andy@voodoo.bitnet.}
}

\vskip.1in
\centerline{\sl Department of Physics}
\centerline{\sl University of California}
\centerline{\sl Santa Barbara, CA 93106-9530}

\bigskip
\centerline{\bf Abstract}

This review
is based on
lectures given at the 1992 Trieste Spring
School on String Theory and Quantum Gravity and at the 1992 TASI Summer
School in Boulder, Colorado.

\Date{9/92}

\newsec{Introduction}

Nearly two decades ago, Hawking \Hawkfirst\
observed that black holes are not black: quantum
mechanical pair production in a gravitational field leads to black hole
evaporation. With hindsight, this result is not really so surprising. It
is simply the gravitational analog of Schwinger pair production in which
one member of the pair escapes to infinity, while the other drops into
the black hole.

Hawking went on, however, to argue for a very surprising conclusion:
eventually the black hole disappears completely, taking with it all
the information carried in by the infalling matter which originally
formed the black hole as well as that carried  in
by the infalling particles created over the
course of the evaporation process. Thus, Hawking
argued, it is impossible
to predict a unique final quantum state for the system. This
argument initiated a vigorous debate in the physics community which
continues to this day.

It is certainly striking that such a simple thought experiment, relying
only on the basic concepts of general relativity and quantum
mechanics, should apparently threaten the deterministic foundations of
physics. We are used to the idea  in quantum mechanics that
we can only predict probabilities of measurements, but the
wave function still has a well defined unitary time evolution.
If Hawking is correct, we would have to give up even this.
This thought experiment brings to mind Einstein's train paradox, together with
the hope that a deeper understanding of the problem will lead to
fundamental new insights.

Unfortunately, subsequent investigations have been unable to definitively
refute or verify Hawking's claim, for several reasons. One is the
problem of the non-renormalizability of quantum gravity inevitably
encountered in the late stages of the evaporation process when the
curvature becomes large. A second difficulty is the inherent complexity
of the process of formation and evaporation of a macroscopic black
hole involving many degrees of freedom.

When a problem cannot be solved, it is often a good idea to consider
a new and related problem which might be solved. An interesting new problem
involves magnetically charged black holes. These can be stabilized
against Hawking radiation in the ``extremal'' limit for which the mass
and charge are equal in Planck units. An incoming particle can excite
such a black hole, which should then decay back to its extremal
state. The new problem is to describe extremal black hole - particle
scattering. Since the scattering involves Hawking emission, there is the
potential for information loss.

In the last year it has been realized that this new problem is far simpler
than that of formation/evaporation of four-dimensional black holes. Indeed,
in an appropriate limit it is equivalent to a model of black hole
formation/evaporation in ${1} + {1}$ dimensions \refs{\CGHS,\BDDO,\dxbh}
and can be fruitfully attacked with the powerful methods of two-dimensional
conformal field theory. In these lectures, we shall review some
basics of black hole evaporation, together with recent results along these
lines.

The outline of this review is as follows.
We begin in section two with a brief review of Penrose diagrams,
apparent horizons, event horizons, extremality and other basic
concepts  used to discuss black holes
in $3+1$ dimensions.
In section three we give a brief overview of black hole
thermodynamics and discuss the puzzles raised by Hawking
radiation, together with some possible resolutions of those puzzles.
In section four the puzzles are discussed in the context of particle-hole
scattering.
In section five we discuss dilaton
gravity and dilatonic black holes, the investigation of
which leads to new insights about black holes in general.
The main emphasis is on
the structure of extremal black holes in this theory
and their relation to the $1+1$ dimensional
theory discussed in the following
sections.
Section six introduces $1+1$ dimensional dilaton gravity
coupled to conformal matter
and discusses the classical solutions of this theory,
including those
that correspond to black holes formed by infalling matter.
The discussion of quantum effects in this theory is begun in
section seven with a presentation of the relation between
Hawking radiation and the trace anomaly and a calculation of
Hawking radiation for a black hole formed by collapse of matter.
In section eight we explain how to include back-reaction
at a semi-classical level, and discuss some of the successes as
well as the limitations of this procedure.
In section nine we discuss constraints on the
full quantum theory and directions for further research.
Brief concluding comments are made in section ten.

This is not meant to be a comprehensive review of
black hole physics. Rather we have attempted to
present an accessible account of selected recent
developments, together with some necessary
background material.

We use Planck units throughout.

\bigskip
\newsec{Black Holes in Einstein Gravity}

We will begin with a brief review of the basic terminology
used to discuss black holes in $3+1$ dimensions.
More details can be found in textbooks such as \wald.
We will present several examples of Penrose diagrams of
increasing complexity, and discuss the notions
of horizons, trapped surfaces, and extremal black holes.

\subsec{Minkowski space}

The line element for Minkoswki space
in spherical coordinates $(t, r, \theta , \phi)$
is given by
\eqn\one{
ds^2 = (-dt^2 + dr^2) + r^2 (d \theta^2 + \sin^2 \theta d \phi^2 ) \equiv
(-dt^2 + dr^2 ) + r^2 d \Omega_{II}^2 .}
At each point $(r,t)$ with $ - \infty < t < \infty , ~ 0 < r < \infty$
there is an $S^2$ of area $4 \pi r^2$.
In what follows we focus on the $(r,t)$ plane and suppress the presence of
the two\--spheres.
It is often useful to introduce light\--cone coordinates
\eqn\two{
\eqalign{
u &= t - r, \cr
v &= t + r, \cr}}
so that $-dt^2 + dr^2 = - du dv$.

\ifig\fone{Relation between $(r+t)$ coordinates and light-cone
coordinates $(u, v)$ and various asymptotic regions of Minkowski space.}
{\epsfysize=4.5in \epsfbox{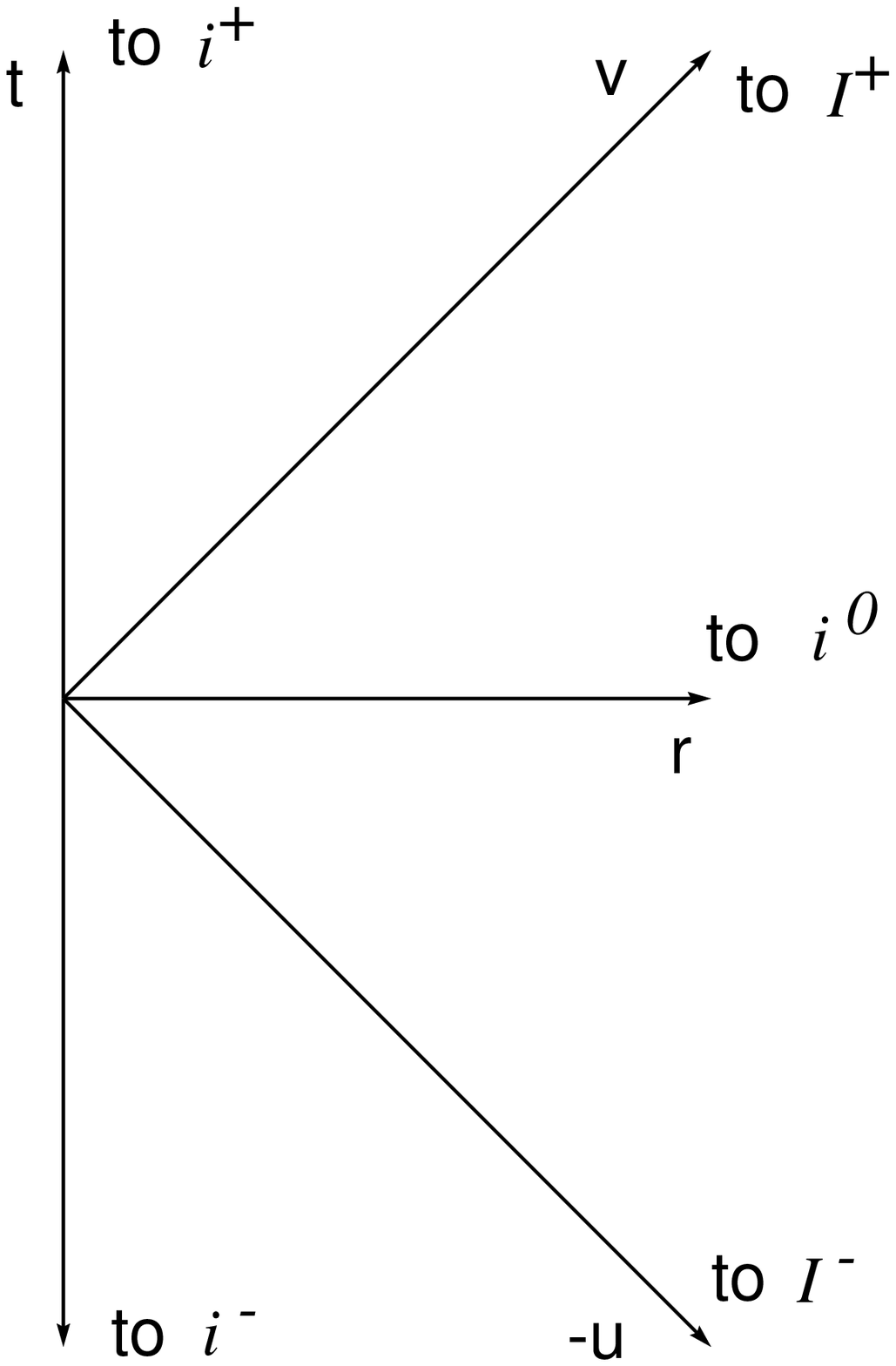}}

The relation between $(r, t)$ and $(u, v)$ and
various asymptotic regions which will play a role in the
following discussion are indicated in \fone.
These are:

\bigskip

$\qquad i^+ = \{ t \air + \infty$ at fixed $r \} =$ future timelike infinity,

\medskip

$\qquad i^- = \{ t \air - \infty$ at fixed $r \} = $ past timelike infinity,

\medskip
$\qquad i^0 = \{ r \air \infty$ at fixed $t \} =$ spacelike infinity,

\medskip
$\qquad \ci^+ = \{ v \air \infty$ at fixed $ u \} =$ future null infinity,

\medskip

$\qquad \ci^- = \{ u \air - \infty$ at fixed $v \} =$ past null infinity.

Future and past null infinity are useful concepts when dealing with radiation.
For example, to measure the mass of an object one needs to know the
deviation of the metric from flat space at large distances.
If the object emits a pulse of radiation at time $t$ and we want to know the
resulting change of mass then, at radius $r$, we must wait a time $t \geq
r$ until the radiation is past to measure the new metric.
As $r \air \infty$, we end up making the measurement at $\ci^+$.

\ifig\ftwo{Relation between $(r+t)$ coordinates and light-cone
coordinates $(u, v)$ and various asymptotic regions of Minkowski space.}
{\epsfysize=4.5in\epsfbox{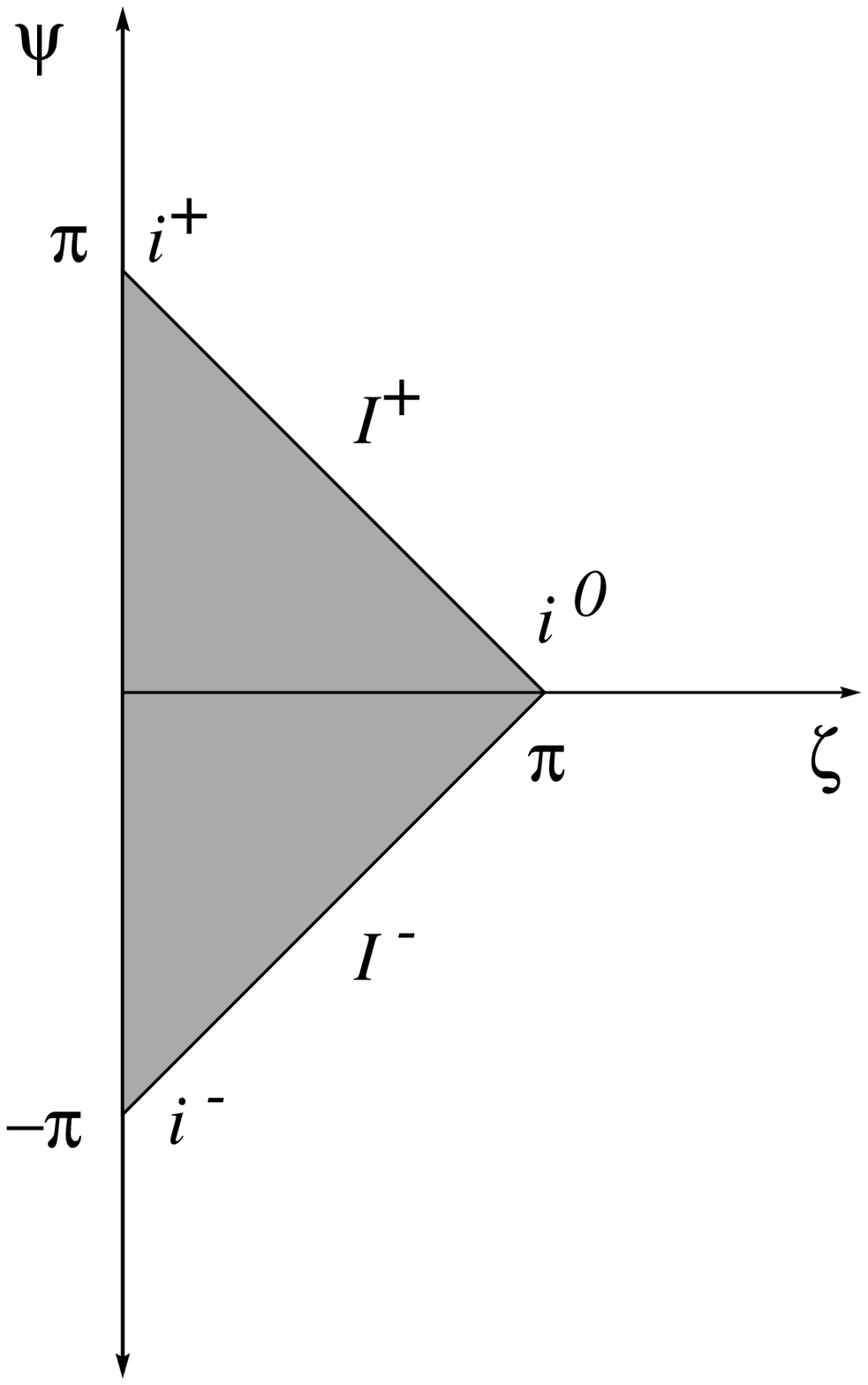}}

A more precise definition of these concepts can be given in terms of
conformal infinity.
We introduce coordinates $( \psi , \zeta )$ with
\eqn\three{\eqalign{
v &= t+r = \tan \frac{1}{2} ( \psi + \zeta ), \cr
u &= t-r = \tan \frac{1}{2} ( \psi - \zeta ), \cr}}
so that
\eqn\four{
ds^2 = \Omega^2 ( \psi, \zeta ) ( -d \psi^2 + d \zeta^2 ) + r^2 ( \psi, \zeta )
d \Omega_{II}^2,
}
with
\eqn\five{
\Omega^{-2} ( \psi, \zeta ) = 4 \cos^2 \frac{1}{2} ( \psi + \zeta )
\cos^2 \frac{1}{2} ( \psi - \zeta ) .
}
The new coordinates
$(\psi,\zeta)$ range over the half-diamond $\zeta \pm \psi < \pi,~~\zeta >0$.
We then introduce an unphysical metric $\bar{g}_{\mu \nu}$ which is
conformal to the actual metric $g_{\mu \nu}$
\eqn\six{
\bar{g}_{\mu \nu} = \Omega^{-2} g_{\mu \nu} .
}
Although distances measured with the $\bar g$ metric differ
(by a possibly infinite factor) from those measured with the
$g$ metric, the causal relation of any two points is the
same in both metrics. Thus the causal structure of the $g$-spacetime is
equivalent to that of the $\bar g$-spacetime.
The unphysical metric $\bar{g}$ is well behaved at the values
of $( \psi, \zeta )$ which correspond to the asymptotic regions of $g$ as
shown in \ftwo .

The Penrose diagram of \ftwo\ brings the previous asymptotic regions
into finite points.
Furthermore, even though $\bar{g}$ is not the physical metric, statements
about the asymptotic behavior of fields in the spacetime with metric $g$
can be translated into simple statements about the behavior of fields at
the finite points corresponding to $i^\pm , i^0 , \ci^\pm$ in the
spacetime with metric $\bar{g}$.
This type of discussion can also be applied to solutions such as the
Schwarzschild metric which have an appropriate notion of asymptotic
flatness.
See \wald\ for further details.

The basic feature of a Penrose diagram is that null geodesics are always
represented by $45^o$ lines. Thus it is easy to discern if two
points are in causal contact, which makes the diagrams very useful.
For example a glance at \ftwo\ reveals that all of Minkowski space is in the
causal past of an observer at $i^+$. The price one pays for this is
that distances are not accurately portrayed: two points finitely separated on a
Penrose diagram may or may not be an infinite geodesic distance apart.

\ifig\fthree{Penrose diagram for $1+1$ dimensional Minkowski space (shaded
region).}{\epsfysize=4.5in\epsfbox{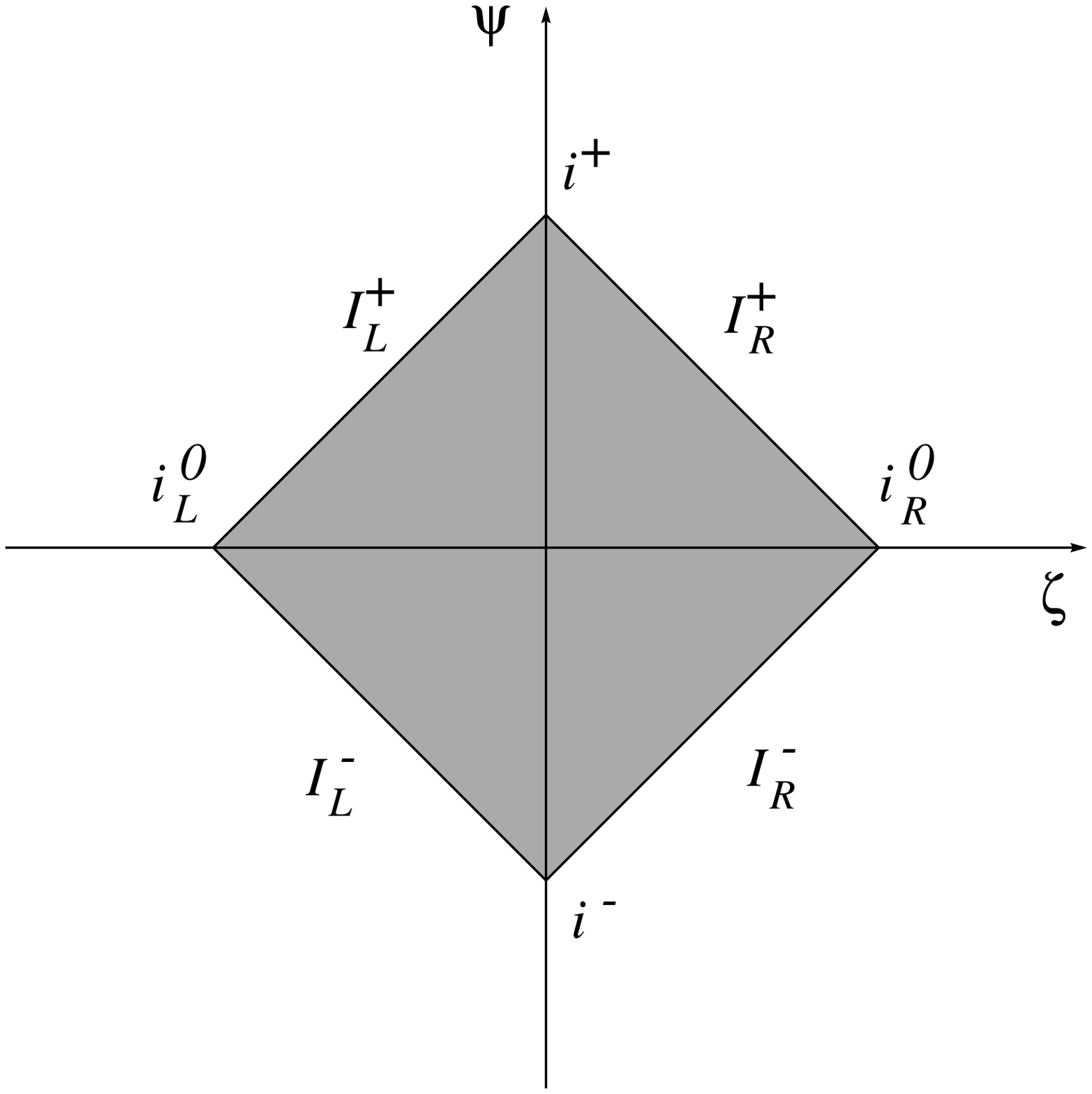}}

\subsec{$1+1$ dimensional Minkowski space}

We have the line element
\eqn\seven{
ds^2 = - dt^2 + dx^2 = - dx^+ dx^-,
}
with $x^\pm = t \pm x$.
Letting
\eqn\eight{
x^\pm = \tan \frac{1}{2} ( \psi \pm \zeta ),
}
where now, since  $- \infty < x <
\infty$, $(\zeta, \psi)$ range over the full diamond $|\zeta \pm \psi|<\pi$.
It follows as in the previous discussion that
the Penrose diagram consists of two copies of \ftwo\ as
shown in \fthree .
There are now two spacelike infinities, $i_{R,L}^0$, corresponding to $x
\air \pm \infty$, and two past and two future null infinities, $\ci_R^\pm
, \ci_L^\pm$ with for example $\ci_R^+$ being where right\--moving light
rays go and $\ci_L^+$ where left\--moving light rays go.
This structure will reappear in section five when we discuss the behavior of
four\--dimensional dilatonic extremal black holes.

\subsec{Schwarzschild Black Hole}

The Schwarzschild black hole with line element
\eqn\nine{
ds^2 = - (1 - \frac{2M}{r}) dt^2 + \frac{dr^2}{(1 - \frac{2M}{r})} + r^2
d \Omega_{II}^2
}
is probably the most familiar non-trivial solution to the vacuum Einstein
equations
$R_{\mu \nu} = 0$.
As is well known, at the origin $r=0$ there is a curvature singularity as
may be verified by calculation of the invariant $R_{\mu \nu \lambda \psi}
R^{\mu \nu \lambda \psi}$.
The singularity in the metric at $r = 2M$ is not a curvature singularity
but instead represents a breakdown of this particular coordinate system.

\ifig\ffour{Maximal analytic extension of the Schwarzschild black hole in
null Kruskal coordinates.}{\hskip.5in\epsfysize=3.6in\epsfbox{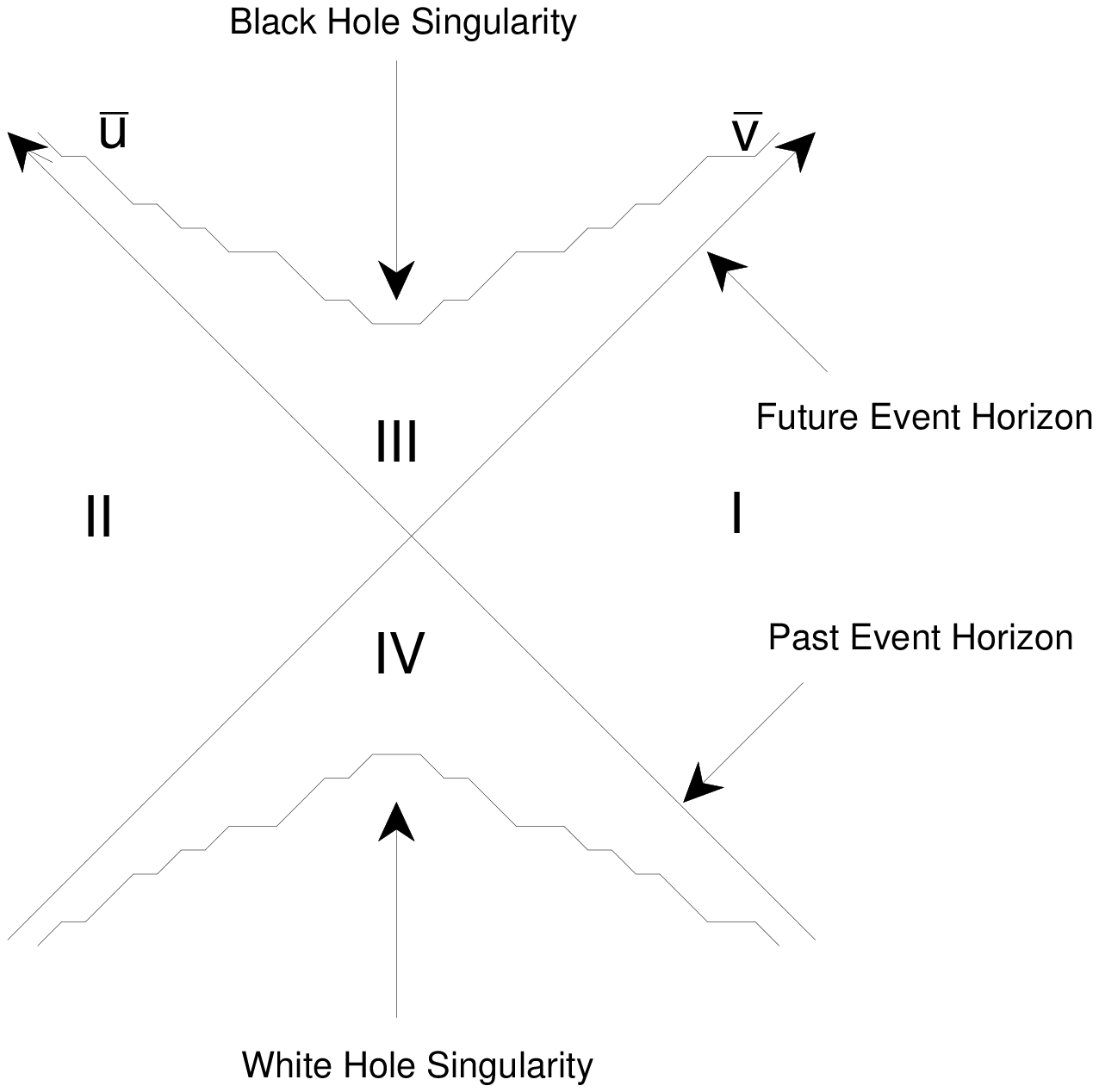}}

The most convenient method to study the behavior near $r = 2M$ is to
introduce coordinates along ingoing and outgoing radial null geodesics.
We thus introduce the tortoise coordinate
\eqn\ten{
r^* = r + 2M \ln (\frac{r}{2M} - 1 )
,}
with $dr = (1 - 2M/r)dr^*$ and
\eqn\eleven{
ds^2 = (1 - \frac{2M}{r}) ( -dt^2 + d{r^*}^2) + r^2 (r^*) d \Omega_{II}^2 .
}
It is clear from \eleven\ that null geodesics correspond to $t = \pm r^*$.
Also note that $r = 2M$ is at $r^* = - \infty$.
The appropriate null coordinates are
\eqn\twelve{\eqalign{
u &= t - r^*, \cr
v &= t + r^* . \cr
}}
The next step is to introduce the null Kruskal coordinates
\eqn\twelve{\eqalign{
\bar{u} &= - 4Me^{-u/4M}, \cr
\bar{v} &= 4Me^{v/4M} . \cr
}}
The region $r \geq 2M$ or $- \infty < r^* < \infty$ maps onto the region
$- \infty < \bar{u} < 0$, $0 < \bar{v} < \infty$.
But now inspection of the metric shows that
\eqn\thirteen{
ds^2 = -\frac{2M}{r} e^{-r/2M} d \bar{u} d \bar{v} + r^2 d \Omega_{II}^2
,}
where $ r (\bar{u}, \bar{v})$ is defined implicitly by \ten\ -- \twelve\ and
it is clear that the metric components are non\--singular at $r = 2M$.
We can thus analytically continue the solution to the whole region $-
\infty < \bar{u} , \bar{v} < \infty$.
The resulting Kruskal diagram of the extension of the Schwarzschild black
hole is shown in \ffour.

\ifig\ffive{Penrose diagram of the analytic extension of
the Schwarzschild black hole.}
{\hskip.25in\epsfysize=2.5in \epsfbox{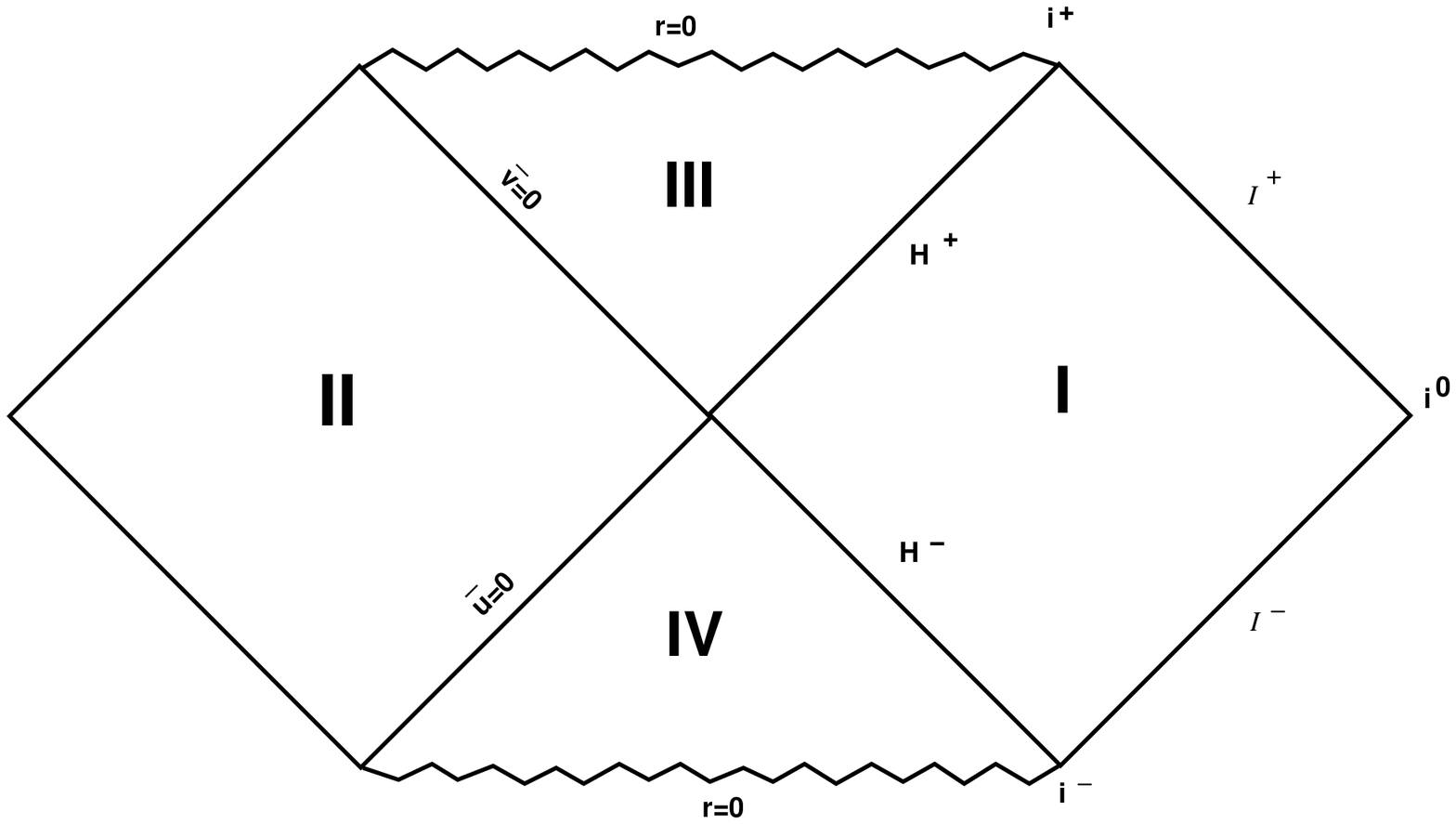}}

A procedure similar to that described earlier for Minkowski space allows
one to bring the asymptotic regions of \ffour\ into finite points in terms
of an unphysical metric $\bar{g}$.
The resulting Penrose diagram for the Schwarzschild black hole is shown
in \ffive .
In this extension of the Schwarzschild metric there are two
asymptotically flat regions denoted I, II in \ffour\ and \ffive .
Also, in addition to the black\--hole singularity (where $r(\bar u, \bar v)$
vanishes)
which reaches $i^+$ in
the infinite future, there is a white\--hole singularity which emerges
from $i^-$ in the infinite past.

\ifig\fstar{Penrose diagram for a black hole
formed by spherically symmetric collapse of radiation.
The solid line is the apparent horizon, which bounds the shaded
region of trapped
surfaces or apparent black hole. The dashed line is the event horizon,
which coincides with the apparent horizon after the collapse is completed.}
{\hskip .5in \epsfysize= 4.8in \epsfbox{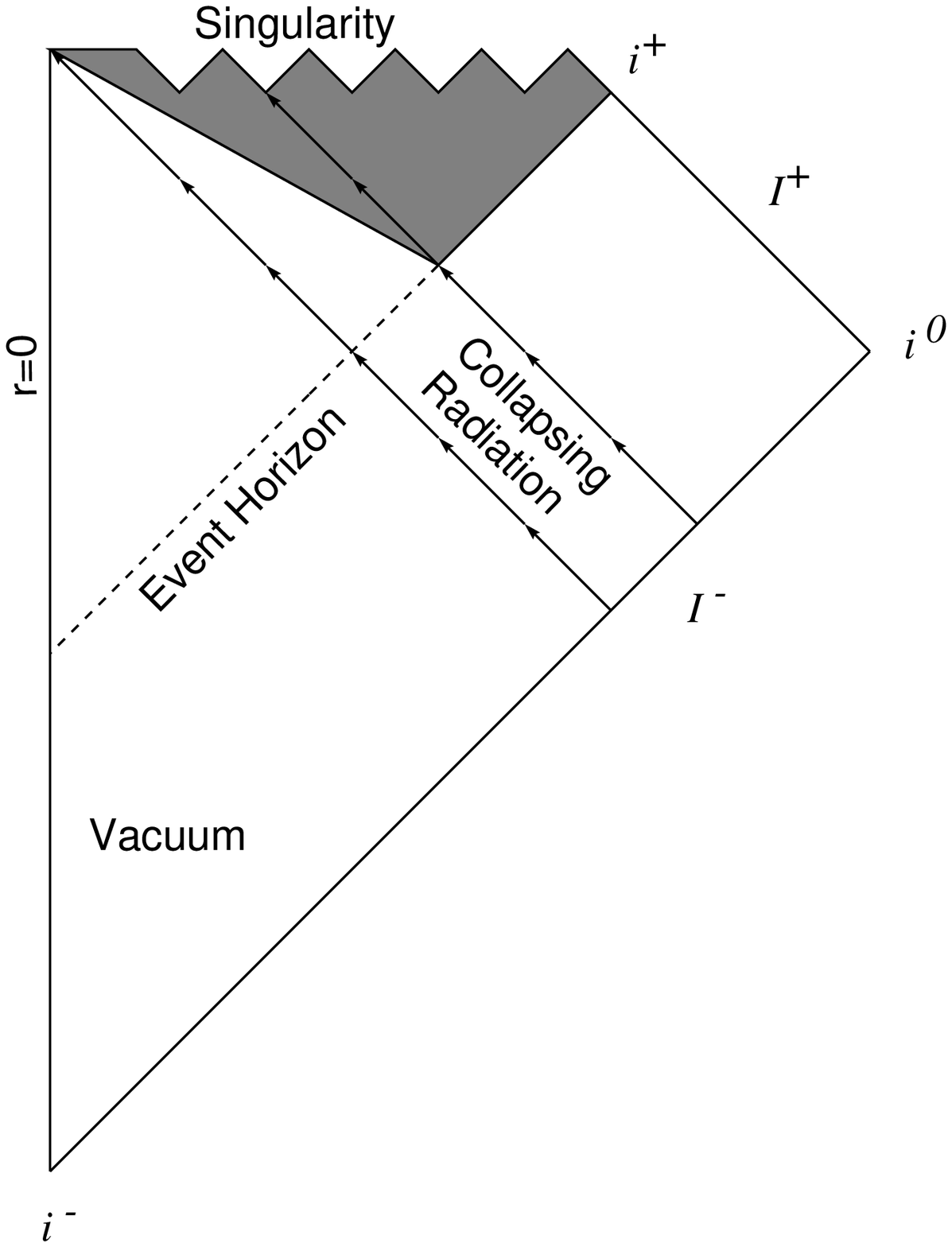}}

It is reasonable to ask how much of this structure is relevant for
classical black holes formed by the collapse of infalling matter.
Only region I and part of region III will exist
for such a physical black hole.
Why this is so should be clear from the Penrose diagram of \fstar\ for a
black hole formed by collapse.

\subsec{Event Horizons, Apparent Horizons and Trapped Surfaces}

In this subsection we will describe the notions of event horizons,
apparent
horizons, and trapped surfaces. We will not give precise definitions for
general surfaces or general spacetimes,
as there
are many subtleties involved. Rather we will attempt to give a flavor of the
ideas in the highly simplified context of spherically symmetric spacetime
geometries and symmetric surfaces.  The statements made in this
section refer only to such surfaces and geometries, although many of them can
be generalized. The reader interested in
precise
statements instead of the general flavor should refer to \wald\ and
\pnrs.

A {\it future event horizon} is the null surface from behind which it is
impossible to escape to ${\cal I}^+$ without exceeding the speed of light. A
{\it past
event horizon} is the time reverse of this: a surface which it is
impossible to get behind starting from ${\cal I}^-$. Schwarzchild contains both
a past and future event horizon as indicated in \ffour, \ffive,
while the spacetime representing a black hole formed by
gravitational collapse contains only a future event horizon.

The interior of a black hole generally contains a region of trapped
surfaces. To illustrate this notion, consider a two-sphere in flat
Minkowski space. There are two families of null geodesics which
emanate from the two-sphere, those that go out and those that go in.
The former diverge, while the latter converge. A {\it trapped surface}
is one for which both families of null geodesics are everywhere
converging, due to gravitational forces. It is easy to check that two-spheres
of constant radius behind
the future horizon in Schwarzchild are trapped. Outgoing null geodesics from
the two-sphere exactly at ${r} = {2M}$ of course generate the
horizon itself, whose area is constant for Schwarzchild. This two-sphere
is therefore marginally trapped.

An {\it apparent horizon} is the outer boundary of a region of trapped
surfaces. We will also find it convenient to refer to a region of
trapped surfaces as an {\it apparent black hole}.

The notions of an apparent horizon and an event horizon are quite different,
although the two are sometimes confused as they happen to coincide for
Schwarzchild. An event horizon is a global concept, and the entire
spacetime must be known before its existence or location can be
determined. The location of an apparent horizon, in contrast, can be
determined from the initial data on a spacelike slice.

To illustrate this, consider a black hole geometry with an apparent
horizon at time ${t_0}$. Throwing matter into the black hole at
a time ${t>}{t_0}$ (relative to any smooth time slicing which goes through the
black
hole) will have no effect on the area or location of the apparent horizon
at time ${t_0}$ (although it will increase its area for later times). However,
the infalling matter does cause the event horizon at the earlier time
${t_0}$ to move out to larger radius. The apparent and event horizons for
a black hole formed by collapsing radiation are illustrated in \fstar .

In classical general relativity, the apparent horizon is typically a null
or spacelike surface which lies inside or coincides with
the event horizon (assuming cosmic censorship) \pnrs.
This is not true when the effects of Hawking radiation are taken into account,
in which case -- as will be illustrated in section three -- the apparent
horizon  can shrink, become timelike and
move outside the event horizon.

It is important to stress that there is no evidence for the existence of
black hole event horizons (as opposed to apparent horizons) in the real
world. In order to answer this question one must follow the apparent
black hole all the way to the endpoint of Hawking evaporation.

\subsec{Kerr-Newman and Extremal Black Holes}

In the early 1970's a number of uniqueness theorems were proved for
gravity coupled to various matter fields.
These results were eventually summarized in a no\--hair ``theorem''
stating that the most general black hole is characterized uniquely by
those quantities which can be measured by the coupling to massless gauge fields
at infinity -- namely by its mass $M$, electric and magnetic charges
$Q_e, Q_m$, and angular momentum $J$. The solution with general
$(M,J)$ was first constructed by Kerr and subsequently generalized by
Newman to arbitrary $(M,J,Q)$ (where $Q^2=Q_e^2+Q_m^2$).
It is worth emphasizing however that there is no single ``no\--hair''
theorem and that many of the specific theorems make assumptions which
rule out physically interesting cases.

There have recently been efforts to understand what kinds of
hair black holes may have in more general theories. For example
at the classical level solutions have been found in which
massless \bartnik\  or massive \monhole\ non-abelian fields
have non\--trivial behavior outside the horizon of the black hole,
although the former are unstable \bizon.
Solutions with ``Skyrmion hair'' were investigated in \luckock.
It has also been pointed out \hair\ that black holes can carry
``quantum hair'': {\it i.e.} charges which are invisible classically but
which in principle can be detected by quantum\--mechanical interference
experiments.
The resulting quantum modifications of black hole structure are
described in \toupee.

A discussion of the general Kerr-Newman solution is beyond the scope of
these lectures.
The special case of $J=0$, first studied by Reissner and Nordstr{\o}m,
will however play some role in
the following.
The Reissner\--Nordstr{\o}m solution is characterized by the
line\--element
\eqn\fourteen{
ds^2 = - (1 - \frac{2M}{r} + \frac{Q^2}{r^2}) dt^2 + (1 -
\frac{2M}{r} + \frac{Q^2}{r^2})^{-1} + r^2 d \Omega_{II}^2.
}
\noindent This solution has a curvature singularity at $r=0$ as for
Schwarzschild,
but in addition has two horizons where $1 - 2M/r +Q^2/r^2$  vanishes:
\eqn\fifteen{
r_\pm = M \pm (M^2 - Q^2)^{1/2} .
}
For $Q < M$ the singularity at $r=0$ is hidden behind these horizons
while for $Q>M$ there are no real roots of \fifteen\ and the solution has a
naked singularity.

The extremal Reissner-Nordst{\o}m solutions with $M=Q$ have special
properties, and will play an important role in the following.
At the classical level, these solutions are just on the verge of
developing a naked singularity. For the special case $Q=0$, the
extremal solution is just the flat space vacuum. We shall see that even for
non-zero $Q$, the extremal solutions can be thought of as the vacuum in
the charge $Q$ sector of the theory. The extremal solutions are also
singled out in the context of supergravity, in that they are solutions of
the supergravity equations of motion which preserve half of the
supersymmetries \gihu.

\newsec{Black Hole Thermodynamics and Hawking Radiation}

During the 1960's and 1970's many fascinating aspects of classical black hole
physics were studied.
One of the main results of these investigations was a set of laws of
black hole mechanics which have a striking analogy with the laws of
thermodynamics, as indicated below.
\medskip
{\settabs 3 \columns
\+ & \it Thermodynamics   & \it Black Hole Mechanics \cr
\+ Zeroth Law & The temperature T is & The surface gravity $\kappa$ is \cr
\+ & uniform over a body in & is constant over the horizon. \cr
\+ & thermal equilibrium.   & \cr
\+ First Law   & $TdS = dE+PdV-\Omega  dJ$ & $\kappa dA = 8 \pi (dM - \Omega
dJ)$ \cr
\+ Second law  & $\Delta S \geq 0$ & $ \Delta A \geq 0$ \cr
}
\medskip
The second law of black hole mechanics relates the change in the
surface area of the horizon, $A$,  for a black hole with angular velocity
$\Omega$ to the changes in the mass and angular momentum and is very
similar in form to the second law of thermodynamics as applied to
rotating systems.
In the above $\kappa$ is
the surface gravity, which can be defined intuitively, for
a static black hole,
as the limiting force which would
have to be exerted at infinity to keep an object stationary
at the horizon.
The third law of thermodynamics that $S \air 0$ as $ T \air 0$, is a
statement about the degeneracy of the ground state of the system and need
not hold for systems with  highly degenerate ground states.

Hawking's remarkable
discovery in 1974 that black holes radiate at a temperature $T =
\kappa /2\pi$ greatly strengthened this analogy, and made it seem very likely
that the laws of black hole physics {\it are} the laws of thermodynamics
as applied to black holes.
This point of view has been further strengthened by all subsequent
investigations. For example Bekenstein's
{\it generalized second law} \bek, which
states that the total
entropy (defined as $S+{1 \over 4}A$ for black hole spacetimes)
always increases, has been verified for processes such as lowering a
box of radiation into a black hole. An excellent recent review can be found in
\bunwa.

This beautiful connection between gravity, quantum mechanics and
thermodynamics is very satisfying, but also raises a number of disturbing
puzzles which must be resolved before we have a full
understanding of this connection.

First of all, the usual laws of thermodynamics can be derived from
microscopic statistical mechanical considerations.
The entropy is calculated as
\eqn\twfive{
S \sim \log ~ ({\rm number~~of~~accessible~~states}),
}
and a careful justification of thermodynamic behavior requires ergodic
behavior.
For a Schwarzschild black hole the entropy is (in Planck units)
\eqn\twsix{
S = A/4 = 4 \pi M^2.
}
This would suggest that a black hole of mass $M$ has of order
$\exp(4 \pi M^2)$ states
and that the black hole is equally likely to be in any of these states.
At present the microscopic description of these states is a mystery, as is
the justification for equal population of these states.

The second major puzzle raised by Hawking's result is that it
raises the possibility
that pure states can evolve into mixed states, resulting
in a net loss of information.
For example one could imagine starting with a pure state of infalling
matter which collapses to form a black hole.
The black hole will then slowly (at first) evaporate. Each pair of particles
produced in the evaporation process is in a pure state. One member
of the pair escapes to infinity. The other member (with negative energy)
falls into the black hole, carrying with it information in the form of
correlations with the outgoing particle. This process continues until the
black hole is Planck-sized, at which point quantum gravity effects are
important and Hawking's semiclassical calculation is no longer expected
to be a good approximation. What happens next is very controversial.
Several possibilities are:

\ifig\floss{Alternative I: the black hole disappears completely, and
information which falls into the singularity is irretrievably lost.
The region of trapped surfaces is shaded .}
{\hskip .5in \epsfysize= 4.8in \epsfbox{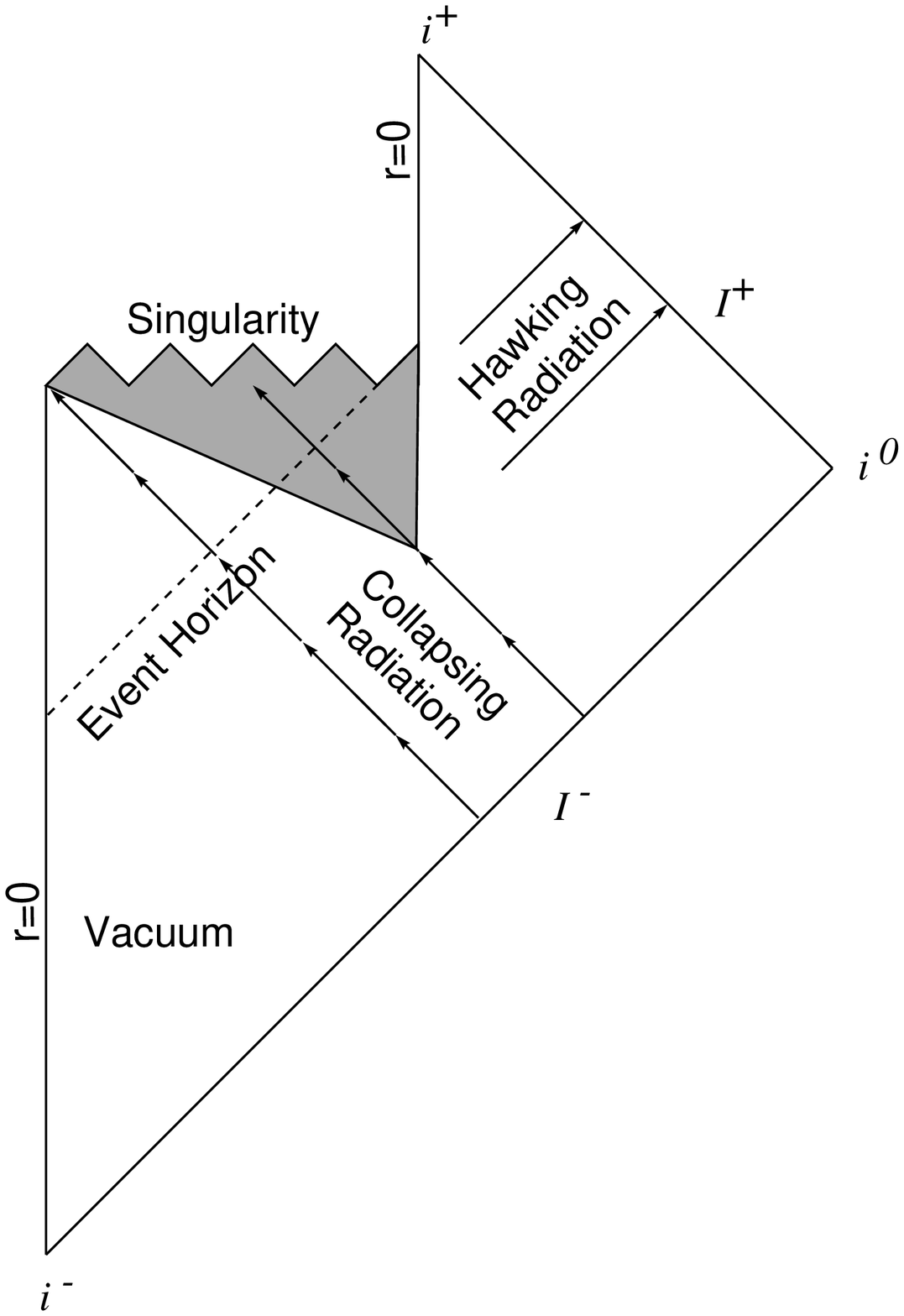}}

\item{(I.)}{{\it The black hole disappears
completely, carrying with it all the
information carried in by the infalling particles.}
This is the alternative advocated by Hawking \Hawktwo~~and is depicted
in \floss.
While not
(as far as we know) logically inconsistent, it is a radical proposal,
implying that the laws of quantum physics are not deterministic. Instead
one can only assign a probability that the outcome of black hole
formation/evaporation is a given quantum state and the rules
for assigning this probability have yet to be fully understood.}

\ifig\frem{Alternative II: Planck-mass remnants. A large apparent black
hole (shaded region) evaporates down to a small remnant, which stores
enough information to ensure that the final quantum state is pure.}
{\hskip .5in \epsfysize=4.8in \epsfbox{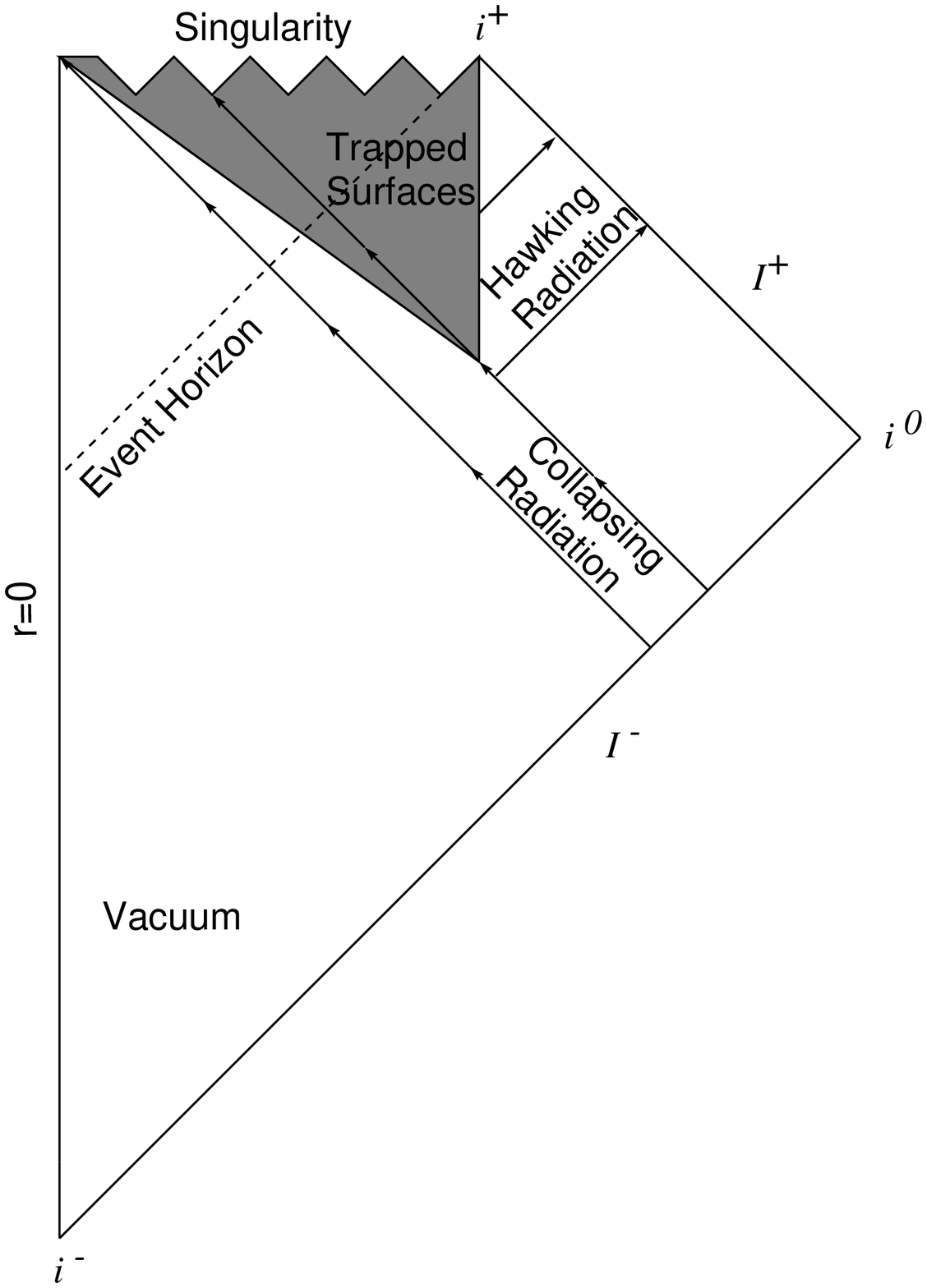}}

\item{(II.)} {{\it
The evaporation process ceases, leaving a stable
Planck-mass remnant. The final state, remnant $+$ radiation is pure.}
This has been advocated for example in \cash. One possible
Penrose diagram for this is depicted in
\frem. Conservation of information requires an
infinite number of ``species'' of stable remnants, one for each
initial state which collapses to a black hole. This leads to
severe phenomenological difficulties. For example there is a
finite -- although incredibly small -- probability for producing a given
species remnant via pair production in a non-static gravitational
field. This probability depends to leading order only on the
coupling of gravity to the remnant -- {\it i.e.}, its mass \foot{Although see
\corn.}. This is roughly the same for every species of Planck mass
remnants. After summing over species, the total remnant production
rate will diverge. Thus it is hard to understand how an infinite number
of remnants could have escaped our attention. }

\ifig\fpure{Alternative III: outgoing radiation carries out all
information contained in the infalling matter. An ``apparent black
hole'' (shaded region) exists for a long time, but a global event horizon
never forms .}
{\hskip .5in \epsfysize=4.8in \epsfbox{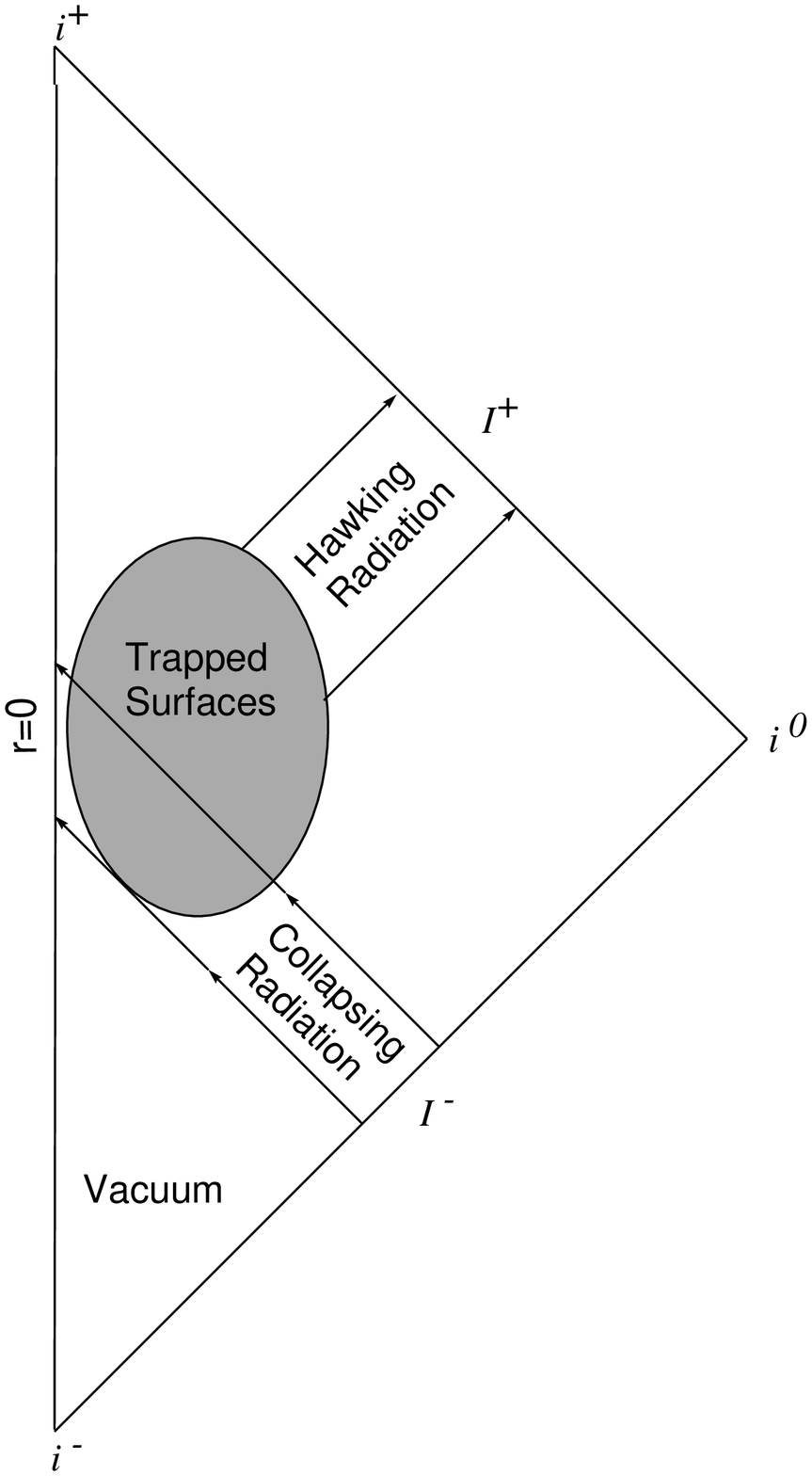}}

\item{(III.)} {{\it The black hole disappears completely.
Outgoing radiation
is correlated with infalling matter and radiation
in such a way that the final pure radiation state is pure.}
This point
of view has been advocated by Page \page, `tHooft \tHooft\ and in \suss\
(to name a few) and is depicted in \fpure.
This is in fact how the final state of {\it e.g.} a burning
lump of coal manages
to be pure. While at each point in the
process the emitted radiation appears thermal, late-time radiation
has subtle correlations with early time radiation. The big difference
for black holes (as stressed in \walddewit)
is that until the final Planckian stage of the
evaporation process they are surrounded by an apparent horizon which
is very nearly null. The infalling particles therefore carry
the information into a region causally shielded from  that part of
future null infinity which precedes (in retarded time) the final
stage of evaporation. Thus the information cannot come back out without
violating
macroscopic causality until the black hole has evaporated down to the Planck
size.
It is conceivable that quantum coherence could be restored by radiation emitted
in the final stage of evaporation which is governed by unknown laws
of quantum gravity. However, since the total available energy
is bounded and small (relative to the initial black hole mass),
this is possible only if the radiation is emitted over an extremely long period
\cash. One then has, for all
practical purposes, a remnant and the objections to alternative (II) are
applicable.  }

\item{(IV.)} {{\it None of the above.}
Given the difficulty with the preceding alternatives, this may well be the
most conservative possibility.}

Further discussion of these and other possibilities can be found in
\refs{\walddewit,\steve}.

\newsec{Extremal Black Hole--Particle Scattering}

Despite intense efforts, it has not so far
been possible to determine the final outcome
of black hole evaporation. One reason is that quantum gravity appears to be
important at late stages in the evaporation process, and one must
therefore confront the problem of non-renormalizability of quantum gravity.
In principle string theory could come to the rescue, but in practice
string technology is not sufficiently developed to handle this
problem.\foot{Except perhaps in ${1}{+}{1}$ dimensions, where the
problem is extremely difficult but perhaps not impossible.} A second obstacle
is the many degrees of freedom involved in the formation/evaporation of a
macroscopic black hole. All of these degrees of freedom must be carefully kept
track of in order to determine whether or not information is lost -- a daunting
task.

In fact it is possible to modify and
boil the problem down to a much simpler problem. In this
simpler problem
both of these obstacles are circumvented, yet the key conceptual puzzles
remain.
The first step in
the boiling down process involves extremal black holes. A key feature of
extremal black holes (including those dilatonic
variants yet to be discussed in
section five) is that the Hawking temperature vanishes. Thus they are typically
quantum mechanically as well as classically stable objects.\foot
{This is not true for extremal electrically charged black holes in a
world -- such as the one we live in -- with ${m}<{e}$ charged particles,
which will lead to ordinary Schwinger pair production in the electric
field of the black hole. In our world, magnetically charged black holes
are possibly examples of the type of quantum mechanically stable objects we
wish to
discuss.}  An ${M}>{Q}$ black hole will tend to Hawking radiate down to
its extremal ${M} = {Q}$ state. Thus the extremal black holes are
quantum ground states of the charge ${Q}$ superselection sector of the
Hilbert space. This view is reinforced by consideration of ${N} = {2}$
supersymmetric versions of the theory, for which it can be shown that they
are exactly stable, supersymmetric, quantum ground states.

Next consider throwing a low energy, neutral particle into an
extremal black hole. This will excite it into a non-extremal
state where, classically, it will remain forever. Quantum
mechanically, however, one expects it to decay via Hawking emission back
to its extremal ground state\foot{We ignore here the interesting possibility
that the incoming particle may cause the extremal black hole to fracture
in to smaller extremal black holes.}. The hope is that the entire process can
be
analyzed in a perturbation expansion about the extremal ground state.

In precise analogy with the four possible outcomes of black hole
formation/evaporation discussed in the preceding, there are the
following
possible outcomes of particle-hole scattering:

\item{(I.)}{ {\it The outcome of the scattering experiment is unpredictable.}
This might be expected if the final outgoing particle has correlations with
a negative energy partner which falls into the black hole.}

\item{(II.)} {{\it The final black hole $+$ particle state is predictable, with
an infinite number of possible black hole states.}
This is the analog of the remnant alternative of the previous
subsection. If the quantum state of the black hole cannot be measured,
this may be indistinguishable  in practice from alternative (I). On the
other hand, if it is possible to measure the quantum state of the black
hole via its long-range quantum hair \refs{\hair,\toupee}\ , or short-range
quantum whiskers \dxbh , this alternative is distinct from (I).
(II) also differs from (I) when one considers the rate of
quantum pair production of black holes, which is proportional to the number of
states.}

\item{(III.)} {{\it The final state of the black hole $+$ particle is
predictable, with a finite number of possible black hole states.}
In this
alternative, most of the (infinite amount) of information in the
incoming particle is carried back out by the outgoing particle (as
in alternative (III) of the preceding section), with possibly a finite
amount transferred to the black hole due to a finite ground state
degeneracy.}

\item{(IV.)}{ {\it None of the above.}
Again perhaps the most likely candidate.}

{}From the preceding discussion it should be clear that the problem of
extremal black hole--particle scattering contains all the same
puzzles as the problem of black hole formation/evaporation. Yet it is clearly
much simpler, as there are few possible final states of the
system, relative to the many possible outcomes of
macroscopic black hole evaporation.

To make real progress, however, we must boil the problem down even more.
We can do this by considering only low-energy scattering
of states with zero angular momentum.
In general one cannot simply truncate
a quantum field theory to the $S$-wave sector and hope to obtain reliable
results. One needs a small parameter to justify the approximation. Such
a parameter is not known for Reissner-Nordst{\o}m black holes. However
in the next section we shall see that the $S$-wave approximation is justified
for extremal dilatonic black holes if a) the black hole is large in
Planck units and b) the Compton wavelength of the incident particle
is large relative to the black hole.
The problem can then be systematically reduced to a ${1} + {1}$ dimensional
quantum field theory living in the $({r}, {t})$ plane of the black hole
(whose properties are the subject of sections six through nine). Clearly the
number of degrees of freedom have been drastically reduced.

It is also important to note that quantum gravity is renormalizable in
two dimensions. Thus the problems of the nonrenormalizability of
four-dimensional quantum gravity are apparently avoided. The reason
behind this is that particle-hole scattering
is essentially a problem in low-energy physics, and the laws of physics
at high energies should not be relevant.\foot{Although it
remains a logical possibility that instabilities inevitably drive one
into the high energy regime.}

\newsec{Black Holes in Dilaton Gravity}

The low-energy limit of string theory with unbroken supersymmetry
generically includes a massless scalar field $\phi$ -- termed the dilaton --
which couples to other fields in a specific way.
In ``realistic'' scenarios with broken supersymmetry it is expected (or
hoped) that the dilaton will acquire a mass which makes its existence
consistent with the standard post\--newtonian tests of general relativity.
Nonetheless, the theories with a massless dilaton should be relevant for
the study of stringy black holes small relative to the Compton wavelength of
the dilaton,
and more generally provide a useful model
for investigating modifications of black hole structure.
The effects of a dilaton mass have recently
been studied in \dmass.

The simplest four\--dimensional example has the dilaton and a $U(1)$ gauge
theory coupled to gravity with action
\eqn\sixteen{
S = \int d^4 x \sqrt{-g} e^{-2 \phi}(R + 4  g^{\mu \nu} \nabla_\mu \phi
\nabla_\nu \phi -\half g^{\mu \lambda} g^{\nu \rho}
 F_{\mu \nu} F_{\lambda \rho}) .
}
In the form \sixteen\ the Einstein term in the action does not have its
canonical form and the dilaton kinetic energy term has the ``wrong'' sign.
This can be remedied by performing a conformal transformation
\eqn\seventeen{
\tilde{g}_{\mu \nu} = e^{- 2 \phi} g_{\mu \nu}.
}
This results in the expression
\eqn\eighteen{
{S} =  \int d^4 x \sqrt{\tilde{g}} (\tilde{R} - 2
\tilde{g}^{\mu \nu} \nabla_\mu \phi \nabla_\nu \phi - \half e^{-2 \phi}
\tilde{g}^{\mu \lambda} \tilde{g}^{\nu \rho} F_{\mu \nu} F_{\lambda \rho}).
}
The metric $\tilde{g}_{\mu \nu}$ is referred to as the canonical metric
(because the Einstein action has the canonical prefactor)
while $g_{\mu \nu}$ is often called the sigma\--model or string metric.
The motivation for this latter terminology is as follows.
The equation of motion for a classical test string in fundamental string
theory implies that its world sheet is a minimal surface with respect
to the metric $g_{\mu\nu}$ -- the
analog of a geodesic for point particles.
Thus it is reasonable to say that strings ``see'' the metric $g_{\mu
\nu}$ rather than $\tilde{g}_{\mu \nu}$.

While we will use mainly the metric $g$, it is important to note that
the stress tensor obtained by varying the matter action with respect to
$g$ does not obey the usual positivity conditions. Therefore the
classic theorems of general relativity -- such as the area and singularity
theorems -- are not applicable in this context. On the other hand, the
stress tensor defined by $\tilde g$ variation is positive,
and the usual theorems do apply. The
area theorem of black hole mechanics will then imply the generalized
second law if
the black hole entropy is
identified as the area in the metric $\tilde g$ (rather than $g$) --
an identification which is confirmed from other points of view \psstw.

Two important points follow trivially from inspection of the
canonical form \eighteen\ of the action.
The first is that for $F_{\mu \nu} = 0$ the solutions reduce to solutions
of Einstein gravity coupled to a massless scalar field.
The no\--hair theorems then imply that the unique black hole solution is
the Kerr solution characterized by $(M, J)$.
The second is that since there is no potential for $\phi$, we are free
to choose an arbitrary constant value $\phi_0$ for $\phi$ at infinity.
{}From \eighteen\ we then see that $e^{ 2 \phi_0} \equiv
g_s^2$ plays the role of the electromagnetic coupling
constant squared at infinity.
In fact, given a solution for one value of $\phi_0$ it is possible to
construct a solution with any other value by utilizing a
classical symmetry of the equations of motion following from the
action \eighteen. If we perform a constant
conformal transformation on the metric accompanied by a constant
shift of $\phi$:
\eqn\oopsone{
{\tilde g}_{\mu \nu} \air e^{-2 \alpha} {\tilde g}_{\mu \nu} , \qquad
 \phi \air \phi+\alpha ,}
then the action \eighteen\ transforms as $S \air e^{-2 \alpha} S$. This
is not a quantum symmetry since the action is not left invariant, but
it is sufficient to insure that the classical equations of motion are
invariant. Therefore one can
generate a new classical solution with a shifted constant
value of $\phi$ at infinity by acting with the transformation \oopsone.

Black holes with non\--zero charge $Q$ have been investigated in
this theory in \gm\ and \ghs\ and are reviewed in \hortr.
The general spherically symmetric solution with $J=0$ and magnetic charge
$Q$ is given in terms of $\tilde{g}_{\mu \nu}$ by
\eqn\tw{\eqalign{
d\tilde s^2 &= \tilde{g}_{\mu \nu} dx^\mu dx^\nu = - (1 - \frac{2M}{r})dt^2 +
\frac{dr^2}{(1-2M/r)} + r ( r - \frac{Q^2}{2M} e^{-2 \phi_0}) d
\Omega_{II}^2 \cr
e^{-2 \phi} &= e^{-2 \phi_0} ( 1 - \frac{Q^2}{2Mr} e^{- 2 \phi_0}), \cr
F &= Q \sin \theta d \theta \wedge d \phi .
}}
Comparing this solution to the Reissner\--Nordst{\o}m solution
described in the previous section one notices several differences.
First of all, this solution has only one horizon at $ r = 2M$ and not two
as for Reissner-Nordst{\o}m.
In fact, the metric components $\tilde{g}_{00}$ and $\tilde{g}_{rr}$ are
exactly  those of the Schwarzschild solution.
Second, the curvature singularity in these variables occurs at $r =
\frac{Q^2}{2M} e^{-2 \phi_0}$ where the area of the two\--spheres
goes to zero.
Finally, the extremal solution occurs at $Q^2 = 4M^2 e^{2 \phi_0}$,
rather than at $Q^2 = M^2$ as for Reissner-Nordst{\o}m black holes.

The supersymmetry of this extremal solution (when embedded in $N\ge 2$
supergravity)
has been  described
in \refs{\gm,\gihu,\renata}. Rotating dilatonic black holes have been
constructed in \gary, dyonic solutions  were found in \ddy, and
higher-dimensional solutions were found in \host.

\ifig\feight{An $S$-wave pulse incident on the spatial geometry of an
extremal dilaton black hole.}
{\hskip .5in\epsfysize=2.6in \epsfbox{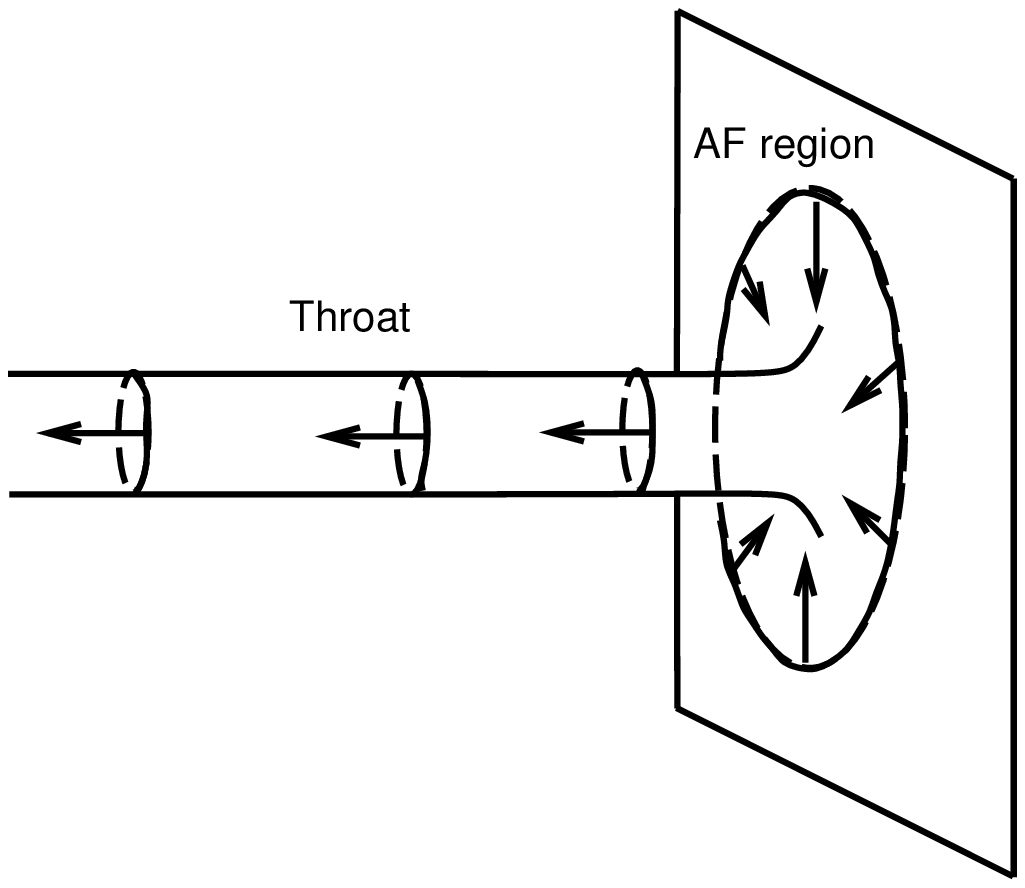}}

Let us look in more detail at the extremal limit of the dilatonic black
hole solution \tw.
As $Q^2 \air 4M^2 e^{2 \phi_0}$ it is clear that both the metric
components and $e^{-2  \phi}$ are becoming singular at the horizon $r =
2M$.
However the physical implications of this are not immediately obvious.
In order to decide whether physical quantities become singular in
theories with scalar fields coupled to gravity we must specify not only
the coupling to the metric but also the coupling to the scalar fields.
In the example at hand the most natural point of view is to ask whether
or not string propagation becomes singular in this limit.
To study this we want to look at the string metric $g_{\mu \nu}$
in the extremal limit.
For $Q^2 \air 4M^2 e^{2 \phi_0}$ the string metric behaves as
\eqn\twtwo{
ds^2 \air e^{2 \phi_0} \left ( - dt^2 +
\frac{dr^2}{(1 - \frac{2M}{r})^2} + r^2 d \Omega_{II}^2 \right ).
}
($r$ and $t$ may be rescaled by a constant factor of $e^{\phi_0}$
to restore the canonical asymptotic behavior of the metric.)
As $r \air \infty$ this approaches flat space with a constant dilaton
$\phi_0$.
As $r \air 2M$ we can introduce a new coordinate $\sigma$ with
\eqn\twthree{
d\sigma = \frac{dr}{1 - 2M/r} =
\frac{rdr}{r-2M} ,
}
so that as $r \air 2M$ we have $\sigma \air 2M \ln (r-2M)$.
We then find that as $\sigma \air - \infty$  ($ r \air 2M$ )
\eqn\twfour{\eqalign{
ds^2 & \air e^{2 \phi_0} ( - dt^2 + d\sigma^2 + (2M)^2 d \Omega_{II}^2 ), \cr
e^{-2 \phi} & \air \frac{e^{-2 \phi_0}}{2M} e^{\sigma/2M}.
}}
We thus see that $\sigma = - \infty $ is an infinite proper distance away and
that as we approach this limit the geometry approaches that of $1+1$
dimensional
Minkowski space $(\sigma,t)$ times a two\--sphere of fixed radius $R =
2Me^{\phi_0} = |Q| $ with a dilaton field $\phi$ linear in the spatial
coordinate $\sigma$ as shown in \feight. The extremal solution has two
asymptotic
regions, one ``out at infinity'' which approaches flat space,
and the other ``down the throat''. Thus the Penrose diagram for this solution
(suppressing the two-spheres of spherical symmetry as usual) is
the same as that of $1+1$ dimensional Minkowski space depicted in \fthree .

Note that in the extremal limit, the geometry is completely non-singular,
and there are no horizons. The black hole is replaced by a
narrow ``bottomless hole'' in space. This non-singular nature of
extremal dilatonic black holes, in contrast to their Reissner-Nordst{\o}m
cousins, simplifies the study of particle-hole scattering.

As an aside we note that a similar simple structure arises in five-dimensional
dilatonic black holes. In this case the black holes are known to be
exact solutions of string theory for some ranges of the parameters
\refs{\chs,\host,\gist}.

It is clear from fig. 10 that low-energy
physics deep into the throat region is effectively two-dimensional,
and is confined to the $({\sigma},{t})$ plane. Excitations in the transverse
directions have energies at least of order ${1}/{R}$. The two-dimensional
effective field theory can be derived by viewing the throat region as a
compactification from four to two dimensions for which the ``internal''
space is a two-sphere (of constant $(\sigma,{t}))$ threaded with magnetic
flux. Standard Kaluza-Klein technology then leads to the effective
two-dimensional action
\eqn\zactn{
{S} = {{1}\over{2\pi}} {\int} {d^2}{x}{\sqrt{-g}} e^{-2 \phi}
({R} + {4} ({\nabla}
{\phi})^2 + {4}{\lambda^2}  -\half {F^2}),}

\noindent where the cosmological constant ${\lambda^2} = 1/4 Q^2$
is a relic of the ``internal''
components of the scalar curvature and field strength
tangent to the compactification two-sphere. Two-dimensional gauge fields
have no local dynamics, and can and will be consistently set to zero
if no charged particles are present. (The role of the $F^2$ term in the
presence of charged fermions is discussed in \malf, charged black holes
in $1+1$ dimensions are discussed in \chargetwod.)

If a low-energy particle is thrown into the non-singular extremal black hole,
it produces a singularity and an  event horizon. This behavior is directly
reflected in
the two-dimensional action \zactn. As shall be seen in the next section, a
particle
incident on the two-dimensional vacuum (which corresponds to the extremal black
hole)
classically produces a singularity and an event horizon.

Before descending to two dimensions there
are several more useful entries in the dictionary relating four and
two-dimensional quantities we would like to explain. In a spherically symmetric
four-dimensional
spacetime,
the area of the two-spheres is given by a function ${4}{\pi}{\tilde R}^2
({\sigma^+}, {\sigma^-})$ where ${\sigma^+}, {\sigma^-}$ are null
coordinates. The two-sphere at ${\sigma^+}, {\sigma^-}$, will be
trapped if ${\tilde R}$  is decreasing in both null directions,
i.e. ${\partial_{\pm}}{\tilde R}<{0}$. The area of the two spheres of
constant ${\sigma^+}, {\sigma^-}$ in the throat region of a near-extremal
geometry as measured in the canonical ${\tilde g}$ metric is
${4}{\pi} Q^2 {e^{-4\phi}}$. Therefore it is natural to define a {\it trapped
point} in the two-dimensional theory as a point at which
\eqn\tpnt{{\partial_{\pm}}{\phi}>{0}.}
\noindent An {\it apparent horizon} is then the outer boundary of such
a region at which ${\partial _+}{\phi} = {0}$ \RST\ (since asymptotically
${\partial_+}{\phi}<{0}$ while ${\partial_-}{\phi}>{0}$.) We will
also use the phrase {\it apparent black hole} to refer to
a region of trapped points. This is distinct from a real black hole,
which is a region from which it is impossible to escape to
${\cal I}_R^+$.

\newsec{Classical 1+1 Dilaton Gravity}
\subsec{Eternal Black Holes}
In this subsection we will be discussing a $1+1$ dimensional theory of gravity
coupled to a dilaton field $\phi$ with action
\eqn\twseven{S_D= {1 \over 2 \pi} \int d^2 x \sqrt{-g} e^{-2 \phi} \left[
  R + 4 (\nabla \phi)^2 + 4 \lambda^2 \right] . }
This action can be viewed as the dimensionally reduced low-energy effective
action for the extremal dilatonic black hole as
described in the previous section. It is also of interest in its
own right as a ``toy'' model of quantum gravity in two dimensions
which, as we shall see, contains black holes and Hawking radiation.
This model arises in two-dimensional non-critical string theory and as such
its black hole solutions were first discovered in \WittTwod\ and \Mandal.
Previous work on two-dimensional black holes which is closely related can be
found
in \Mann, and on models of two-dimensional gravity with scalars in
\refs{\rom,\tei, \cham}.

The classical equations of motion which follow from \twseven\ are
\eqn\tweight{
2 e^{-2 \phi} \left[ \nabla_\mu \nabla_\nu \phi + g_{\mu \nu}
( (\nabla \phi)^2 - \nabla^2 \phi - \lambda^2 ) \right] =0,}
\eqn\twnine{ e^{-2 \phi} \left[
R + 4 \lambda^2 + 4 \nabla^2 \phi - 4 (\nabla \phi)^2  \right] = 0,}
where the first equation results from variation of the metric
and the second is the dilaton equation of motion.  We first
note that there is a solution (often called the linear dilaton
vacuum) characterized by
\eqn\thirty{
R = \nabla^2 \phi = 0 , \qquad (\nabla \phi)^2 = \lambda^2 .}
We shall refer to this simply as the vacuum.
We can introduce coordinates $(\sigma , \tau)$ so that
\eqn\thone{
g_{\mu \nu} = \eta_{\mu \nu} , \qquad \phi = -\lambda \sigma,}
in the vacuum. Note that the vacuum is not translationally invariant -- a
feature
which also occurred in the ``throat'' region of the
extremal dilaton black hole. As was the case there, the natural
coupling constant in this theory is $g_s = e^{\phi}$. Thus
the vacuum can be divided into a strong coupling region ( $\sigma \air -
\infty$)
and a weak coupling region ($\sigma \air + \infty$). It is sometimes
useful to think of the strength of the coupling as providing a coordinate
invariant notion of one's location in this one-dimensional world.

To introduce the black hole solution of this theory it is useful
to introduce light-cone coordinates (the relation of these coordinates
to the previous ones will be discussed momentarily)
\eqn\thtwo{
x^\pm = x^0 \pm x^1 ,}
and to choose conformal gauge $g_{\mu \nu} = e^{2\rho}\eta_{\mu \nu}$,  or
in light-cone coordinates
\eqn\ththree{
g_{+-} = -{1 \over 2} e^{2 \rho} , \qquad g_{++} = g_{--} = 0.}
We then have $R=8 e^{-2 \rho} \partial_+ \partial_- \rho$ and the
equations of motion become
\eqn\thfour{\eqalign{
\phi &: \qquad  e^{-2(\phi+\rho)} \left[ -4 \p+ \pmi \phi +
 4 \p+ \phi \pmi \phi + 2
        \p+ \pmi \rho + \lambda^2 e^{2 \rho} \right] = 0, \cr
\rho &: \qquad  e^{-2 \phi} \left[ 2 \p+ \pmi \phi - 4 \p+ \phi \pmi \phi
                     - \lambda^2 e^{2 \rho} \right] =0. \cr }}
Note that these two equations imply
\eqn\thfive{
\p+ \pmi (\rho -\phi) = 0, }
so that $(\rho - \phi)$ is a free field. Since we have gauge fixed
$g_{++}$ and $g_{--}$ to zero we must also impose their
equations of motion as constraints. This gives
\eqn\thsix{\eqalign{
e^{-2 \phi} ( 4 \p+ \rho \p+ \phi - 2 {\p+}^2 \phi ) &= 0, \cr
e^{-2 \phi}( 4 \pmi \rho \pmi \phi - 2 {\pmi}^2 \phi ) &=0 . \cr }}

Now \thfive\ implies that $\rho$ and $\phi$ are equal up to the sum of
a function purely of $x^+$, $f_+(x^+)$ and a function purely of $x^-$,
$f_-(x^-)$. But
a coordinate transformation $ x^\pm \air \tilde x^\pm (x^\pm)$
preserves the conformal gauge \ththree\ and can be used to set $f_\pm=0$.
Thus we can choose $\rho=\phi$ in analyzing the equations of motion.
With this choice the remaining equations and constraints reduce to
\eqn\thseven{\eqalign{
& \pmi \p+ (e^{-2 \rho}) = -\lambda^2, \cr
& {\p+}^2 (e^{-2 \rho}) = {\pmi}^2 (e^{-2 \rho}) = 0, }}
which has the general solution (up to constant shifts of $x^\pm$)
\eqn\theight{
e^{-2 \phi} = e^{-2 \rho} = {M \over \lambda} - \lambda^2 x^+ x^-, }
where $M$ is a free parameter which will turn out to be the
mass of the black hole.

Calculating the curvature we find
\eqn\thnine{
R = 8e^{-2 \rho} \p+ \pmi \rho = {4 M \lambda \over M/\lambda - \lambda^2
                                  x^+ x^-} ,}
which is divergent at $x^+ x^- = M/\lambda^3$. This solution  has the same
qualitative features
as the $(r,t)$ plane of the Schwarzschild black hole. The Penrose
diagram is in fact the same as that in fig. 5 with $(\bar u , \bar v)$
replaced by $(x^- , x^+)$.

Region I in fig. 5 should asymptotically approach the flat space
vacuum. To see this we can introduce coordinates
\eqn\forty{\eqalign{
\lambda x^+ &= e^{\lambda \sigma^+} ,\cr
\lambda x^- &= - e^{-\lambda \sigma^-}. \cr }}
Note that the range $- \infty < \sigma^+ , \sigma^- < +\infty$ covers
only region I of fig. 5. It is also important ro remember that in these
coordinates $\rho$ will no longer equal $\phi$ since $\phi$ transforms as
a scalar under coordinate transformation while $\rho$ does not. In these
coordinates we find that
as $\sigma = (\sigma^+ - \sigma^-)/2 \air \infty$
\eqn\foone{\eqalign{
\phi & \air -\lambda \sigma - {M \over 2 \lambda} e^{-2 \lambda \sigma}, \cr
\rho & \air 0 - {M \over 2 \lambda} e^{-2 \lambda \sigma}, \cr }}
and the solution approaches the vacuum up to exponentially small corrections.
It is also important to note that $g_s = e^{\phi} \air 0$ as
$\sigma \air \infty$ and that at the horizon $x^- =0$, $g_s =
\sqrt{\lambda/M}$. Thus we are in weak coupling throughout region
I for sufficiently massive black holes ($ M >> \lambda$).
\subsec{Coupling to Conformal Matter}
So far all we have constructed is an ``eternal'' black hole
solution. To determine whether such solutions form from
non-singular initial conditions and to study Hawking radiation
we must couple in some dynamical matter degrees of freedom.
%
%
To study this process in our $1+1$ dimensional model we modify
\twseven\ by adding a matter term of the form
\eqn\fotwo{
S_M = -{1 \over 4 \pi} \sum_{i=1}^N \int d^2 x \sqrt{-g} (\nabla f_i)^2, }
where the $f_i$ are a set of $N$ massless matter fields
\foot{Such fields would arise in the reduction
from four dimensions discussed in the previous section if
one began with massless four-dimensional scalars, or after
bosonization, with charged fermions.}.
For the moment we take $N=1$ and will consider general N
when we discuss Hawking radiation and back reaction.
In conformal gauge the $f$ equation of motion is simply
\eqn\fothree{
\p+ \pmi f =0. }

\ifig\ften{Penrose diagram for formation of a black hole
by an $f$ shock-wave.}{\hskip .5in \epsfysize=3.5in \epsfbox{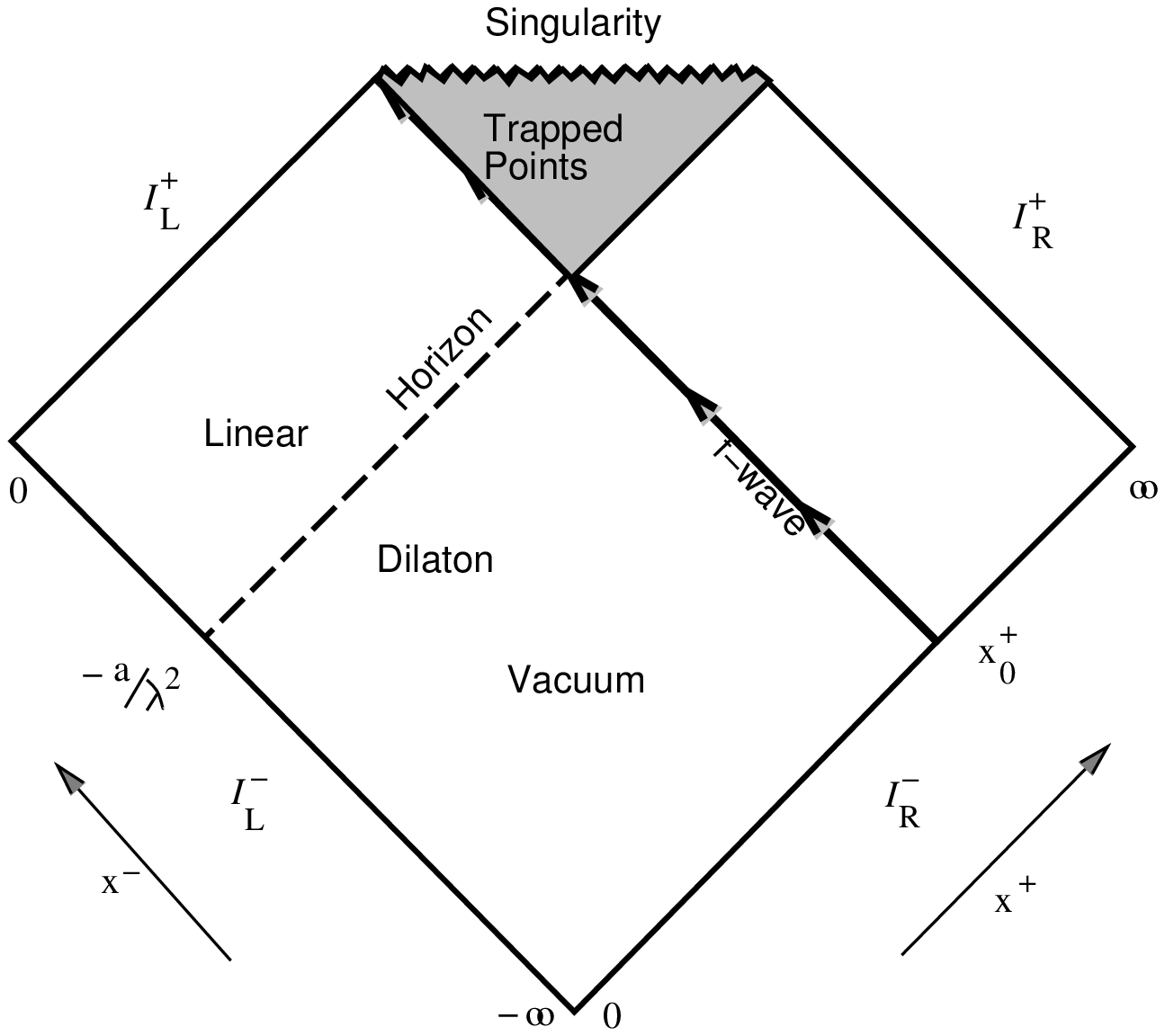}}

Let us consider sending in a pulse of energy from the
right. Although we could consider taking $f$ to be
some function of $x^+$ with finite width, to simplify
the calculation we take the $f$ pulse to be a
shock-wave traveling in the $x^-$ direction with magnitude $a$
described by the stress tensor
\eqn\fofour
{\half \partial_+ f\partial_+ f= a\delta (x^+ - x^+_0)\ .}
The only modification in the equations of motion and
constraints due to the matter fields in this case
is in the $g_{++}$ constraint which becomes
\eqn\fofive{
e^{-2 \phi} ( 4 \p+ \rho \p+ \phi - 2 {\p+}^2 \phi) = -{ 1 \over 2}
             \p+ f \p+ f.}
For $x^+ < x_0^+$ we assume we are in the vacuum, while for $x^+ > x_0^+$ we
know that the solution must be of the form
\theight. Matching the discontinuity across $x_0^+$ we obtain
the solution
\eqn\fosix{ e^{-2\rho} = e^{-2\phi} =-a(x^+ - x^+_0)
\Theta(x^+ - x^+_0) -\lambda^2 x^+ x^- . }
For $x^+ > x^+_0$ this is identical to a black hole of mass $ax^+_0 \lambda$
after
shifting $x^-$ by $a/\lambda^2$.
The Penrose diagram for this
spacetime is shown in \ften.

Both of these classical solutions have a straightforward
interpretation from the four-dimensional
viewpoint. The vacuum corresponds to an extremal $Q^2= 1/(4 \lambda^2)$
black hole, while the two-dimensional black hole solutions
correspond to non-extremal four-dimensional black holes.
The fact that an arbitrarily low-energy $f$-particle incident on the vacuum
produces a two-dimensional black hole corresponds to the
fact that an arbitrarily small particle thrown in to an extremal
black hole turns it into a four-dimensional non-extremal black hole.
Just as one expects this non-extremal black hole to revert to
its extremal state via Hawking emission, one expects the two
dimensional black hole to evaporate.
Thus the four-dimensional process of particle-hole
scattering corresponds to two-dimensional black hole
formation/evaporation.

\newsec{Hawking Radiation and the Trace Anomaly}

So far we have achieved a satisfying description of the classical
formation of a $1+1$-dimensional black hole from collapsing matter. However
the real motivation for
studying this model is to understand quantum effects. We will do this
in several parts. To begin with we will analyze the quantum effects
of matter fields in the fixed classical background of a black hole formed by
collapsing matter.

In two dimensions there is a beautiful relation between the trace
anomaly and Hawking radiation discovered in
\CrFu. For a massless scalar field
the trace of the energy-momentum tensor is zero classically,
$T \equiv T^\mu_\mu = 0$. Quantum mechanically there is a one-loop anomaly
which relates the expectation value of the
trace of the energy-momentum tensor to the
Ricci scalar
\eqn\foseven{\langle T \rangle= {c \over 24} R,}
where $c=1$ for a massless scalar and $c=1/2$ for a Majorana fermion.
In conformal gauge with $T= -4 e^{-2 \rho} T_{+-}$ this gives for
$N$ $c=1$ scalars
\eqn\foeight{ \langle T_{+-}^f \rangle = - {N \over 12} \p+ \pmi \rho .}
Given the expectation value of $T_{+-}$ as above we can use energy-momentum
conservation to determine $T_{++}$ and $T_{--}$. We have
\eqn\fonine{
\p+ T_{--}+\pmi T_{+-}-\Gamma^-_{--}T_{+-}	      =0,}
and similarly for $T_{++}$.
Using $\Gamma^+_{++}= 2 \p+ \rho$, $\Gamma^-_{--} = 2 \pmi \rho$
the solution is found as
%
\eqn\fifty
{\eqalign{\vev{T^f_{++}}& = -{N \over 12} \left(
 \partial_+\rho \partial_+\rho - \partial^2_+\rho +
t_+(\sigma^+)\right)\ ,\cr
          \vev{T^f_{--}}& =-{N \over 12} \left(
  \partial_-\rho\partial_-\rho - \partial^2_-\rho +
t_-(\sigma^-)\right)\ .\cr}}
The functions of integration $t_\pm$  are not determined
purely by energy-momentum conservation and must be fixed by
imposing physical boundary conditions.
For the collapsing
$f$-wave, $T^f$ should vanish identically in the linear dilaton region, and
there should be no incoming radiation along ${\cal I}^-_R$ except for the
classical $f$-wave at $\sigma^+_0$.

We now
 turn to a calculation of Hawking radiation from a ``physical''
black hole formed by collapse of an infalling $f$ shock-wave as in
\fofour.
The calculation and its physical interpretation is clearest in coordinates
where the metric is asymptotically constant on ${\cal I}_R^{\pm}$.
We thus set
\eqn\fione
{\eqalign{e^{\lambda\sigma^+}& =\lambda x^+, \cr
          e^{-\lambda\sigma^-} &= -\lambda x^- - {a\over\lambda}.\cr}}
This preserves the conformal gauge \two\ and gives for the new
metric
\eqn\fitwo{-2 g_{+-} = e^{2 \rho} = \cases{[1+{a \over \lambda} e^{\lambda
\sigma^-}]^{-1},
             & if $\sigma^+ < \sigma_0^+$; \cr
            [1+ {a \over \lambda} e^{\lambda (\sigma^- -\sigma^+ +\sigma_0^+)}
]^{-1}
             & if $\sigma^+ > \sigma_0^+$ \cr}}
with $\lambda x_0^+ = e^{\lambda \sigma_0^+}$.

The formula for $\rho$, together with the boundary conditions
on $T^f$ at ${\cal I}^-_{L,R}$ then implies
\eqn\fifthree
{t_+ = 0, \qquad t_- = {-\lambda^2 \over 4} [1- (1+a e^{\lambda
\sigma^-}/\lambda )^{-2} ]. }
The stress tensor is now completely determined, and one can read off
its values on ${\cal I}^+_R$ by taking the limit $\sigma^+\to \infty$:
\eqn\fiffour
{\eqalign{ \vev{T^f_{++}} &\to 0, \qquad\vev{T^f_{+-} }\to 0,\cr
          \vev{T^f_{--}} &\to { N \lambda^2 \over48} \left[
1-{1\over\left(1+a e^{\lambda\sigma^-}/\lambda\right)^2}\right]~~.}}
The limiting value of $T^f_{--}$ is the flux of $f$-particle energy across
${\cal I}^+_R$. In the far past of ${\cal I}^+_R$ $(\sigma^- \to-\infty)$ this
flux vanishes exponentially while, as the horizon is approached, it
approaches the constant value $N \lambda^2/48$.
This is nothing but Hawking radiation. The surprising result
that the Hawking radiation rate is asymptotically
independent of mass has been found in
other studies of two-dimensional gravity.

Although we have established that there is a net flux of energy
which starts at zero and builds up to a constant value (ignoring
backreaction) the skeptical reader might wonder whether this
is in fact thermal Hawking radiation. There are two ways of arguing
that the radiation is indeed thermal. The first involves a by now
standard trick of rotating the black hole solution to Euclidean
space. One then finds a non-singular solution only if the time coordinate
is periodically identified. Once one does this, Green's functions
constructed in this background will necessarily be thermal and it
is easy to check from the periodicity that the temperature is
$\lambda/2 \pi$.

A more satisfying answer involves canonical
quantization of the matter $f$ fields in the black hole spacetime.
Concentrating only on the right-moving modes which carry the Hawking
radiation, one first considers the asymptotically flat ``in''
and ``out'' regions $\ci_L^-$ and $\ci_R^+$. For the ``in'' region
one has a complete set of states for the mode expansion
of the fields, but for the ``out'' region
one must add a set of modes for the region internal to the black hole
to obtain a complete set. At the end of the calculation these internal
states are traced over since they are not observable. Although
each asymptotic region has a natural timelike coordinate which allows
one to define particles and anti-particles, the definitions do not agree
so that positive frequency modes in one region will be a combination
of positive and negative frequency modes in the other region. This
has the interpretation of particle creation. In particular, the ``in''
vacuum corresponds to a thermal distribution of particles in
the ``out'' region with temperature $\lambda/2 \pi$.
The details of this procedure in this model
can be found in \GiNe. For general background on canonical
quantization in curved spacetimes see \bd.

\newsec{Including the Back-Reaction}

If expression  \fiffour\ is integrated along all of ${\cal I}^+_R$
to obtain the total energy emitted in Hawking radiation an infinite
answer is obtained. This is obviously nonsense:  the black hole can not
radiate more energy than it owns.

The reason for this nonsensical result is simple: the backreaction of
the Hawking radiation on the geometry has been neglected. While this
should be unimportant at early times when the Hawking radiation is weak,
ultimately it should be important enough to terminate the radiation
process when the mass reaches zero.

\subsec{The One Loop Action}

The backreaction is easily included by simply letting the quantum stress tensor
\foeight, \fifty\ act as a source for the classical metric equations. For
example the ${\rho}$ equation \thfour\ is modified to read

\eqn\mrho{{e^{-2\phi}}({2}{\partial_+}{\partial_-}{\phi} - {4}{\partial_+}
{\phi}{\partial_-}{\phi} {-} {\lambda^2} {e^{2\rho}}) = {{N}\over{12}}
{\partial_+}{\partial_-}{\rho},}

\noindent while the constraint equations are modified by the addition
of \fifty. These modified equations can be derived from the non-local
action \poly
\eqn\plact{{S}_D-{{N}\over{96\pi}} {\int} {d^2}{x} {\sqrt{-g}} {R}
{\sq}^{-1}{R},}
\noindent where ${\sq^{-1}}$ is the scalar Greens function. Note that
in conformal gauge ${\sq^{-1}}{R} = -2{\rho}$, so that \plact~~is local.

There is another, equivalent, method of deriving the extra term in \plact.
The quantum theory is defined by the functional integral in conformal
gauge
\eqn\zzz{{Z} = {\int}{\cal D}({b},{c},{\rho},{\phi}){\CD}{f_i}
{e^{{i}({S}_D{+}{S_{bc}}+{S_M})}},}
\noindent where ${b}$ and ${c}$ are Fadeev-Popov ghosts arising from gauge
fixing to conformal gauge, and ${S_{bc}}$ is their action. In order to define
the measures in ${Z}$ one must introduce a short distance regulator. This
should be done in a covariant manner, which implies that the measures
will depend on ${\rho}$ and so should be denoted e.g. ${\CD}_{\rho}{f_i}$.
This dependence of the measure on ${\rho}$ is given by
\eqn\ddd{{\CD}_{\rho}{f_i} = {\CD}_0{f_i}{e^{{-}{{iN}\over{12\pi}}
{\int}{\partial_+}{\rho}{\partial_-}{\rho}}},}
\noindent where ${\CD}_0$ is the measure with ${\rho} = {0}$. The
term in the exponent is precisely the conformal gauge version of the
extra term in \plact. Thus we see that this extra term arises from
the metric dependence of the functional measure.

Of course, there is also metric dependence hidden in the $({b}, {c},
{\rho}, {\phi})$ measure. One might suspect that this simply leads to a
shift of ${N}$ in \plact~~but this is not necessarily the case \Stro. To see
why not,
note that the functional
measure is canonically
derived from the metric governing small fluctuations of the fields.
This is determined from the kinetic part of the action as \refs{\hver,\qtdg}
:
\eqn\sss{{d{\cal S}^2} = {{1}\over{4 \pi}} {\int} {d^2}{\sigma}{\sqrt{-g}}(
{\sum_{i=1}^{N}}{\delta_i}{f}{\delta_i}{f} -8 {e^{-2\phi}}
{\delta}{\phi}{\delta}{\phi} +8 {e^{-2\phi}}{\delta}{\phi}
{\delta}{\rho}).}
\noindent Thus in computing the ${\rho}, {\phi}$ (as well as ${b}, {c}$)
measures it is natural to replace ${g}$ by ${e^{-2\phi}}{g}$, or equivalently
${\rho}$ by ${\rho} - {\phi}$. One then finds \refs{\CGHS,\Stro}\
\eqn\zfin{{Z} = {\int}{\CD}_0 ({b}, {c}, {\rho}, {\phi}, {f_i}) {e^{iS_1}},}
\noindent where
\eqn\sfin{\eqalign{
{S_1}&= {{1}\over{\pi}} {\int} {d^2}{\sigma}
\biggl[{e^{-2\phi}}({2}{\partial_+}
{\partial_-}{\rho} - {4}{\partial_+}{\phi}{\partial_-}{\phi}+ {\lambda^2}
{e^{2\rho}})\cr
&{-} {{N}\over{12}}\,{\partial_+}{\rho}{\partial_-}{\rho} + {{1}\over{2}}
{\sum_{i=1}^N} {\partial_+}{f_i}{\partial_-}{f_i} + {2}{\partial_+}
({\rho}-{\phi}){\partial_-}({\rho}-{\phi})\biggr],\cr}}
\noindent and now all ${\rho}$ dependence has been explicitly exhibited.
The corresponding modifications of the constraints will be given below.

The difference in the ghost-gravity and matter measures has a physically
sensible consequence: there is no Hawking radiation of ghosts or gravity
modes for large black holes. This can be seen by going to a gauge
in which ${\rho}-{\phi}$ vanishes to leading order.
This is as expected because the
dilaton and metric are non-dynamical, while ghosts are not real
particles.

\subsec
{\it The Large ${N}$ Approximation}

The quantum theory defined by \zfin~~is still rather complicated. The theory
simplifies in the limit ${N}{\rightarrow}{\infty}$, with ${N}{e^{2\phi}}$
held fixed \CGHS . The first three terms in \sfin~~are then of order ${N}$,
while the last ghost-gravity term is order ${N^0}$ and therefore (formally)
negligible. Furthermore, since the entire action is large the stationary
phase approximation is valid, so we need merely solve the semiclassical
equations. The semiclassical ${\rho,\phi}$ equations can be cast in
the form
\eqn\phin{8P{\partial_+}{\partial_-}{\phi} = {-}{P^{\prime}} ({4}
{\partial_+}{\phi}{\partial_-}{\phi}+{\lambda^2}{e^{2\rho}}),}
\eqn\rhon{2P{\partial_+}{\partial_-}{\rho} = {e^{-4\phi}}({4}{\partial_+}
{\phi} {\partial_-}{\phi} + {\lambda^2}{e^{2\rho}}),}
\noindent where
\eqn\pn{{P} = {e^{-4\phi}} - {{N}\over{12}} {e^{-2\phi}},}
\eqn\ppn{{P^{\prime}} = {-} {4}{e^{-4\phi}} + {{N}\over{6}} {e^{-2\phi}}.}
\noindent The ${++}$ constraint equation is
\eqn\consn{\eqalign{{T_{++}}&=
{e^{-2\phi}}({4}{\partial_+}{\phi}{\partial_+}{\rho}
- {2}{\partial_+^2}{\phi}) + {{1}\over{2}} {\sum_{i=1}^{N}}{\partial_+}
{f_i}{\partial_+}{f_i}\cr
&{-}{{N}\over{12}} ({\partial_+}{\rho}{\partial_+}{\rho} -
{\partial_+^2}{\rho}) + {t_+} = {0},\cr}}
\noindent while a similar equation holds for ${T_{--}}$.

An immediately obvious feature of \phin\ and \rhon\ is
\refs{\RST,\BDDO}\ that the
prefactor ${P}$ on the left hand side vanishes at the critical value
of the dilaton field:
\eqn\pher{{\phi_{cr}} = {{1}\over{2}} {\ell}{n} {{12}\over{N}}.}
\noindent Unless the right hand sides of \phin\ and \rhon\ vanish when ${\phi}$
reaches ${\phi_{cr}}$ the second derivatives of these fields will have to
diverge. While the RHS of \phin\ and \rhon\ do vanish for the vacuum,
this will not be the case for perturbations of the vacuum, and singularities
will occur.

These singularities can be viewed as a quantum version of the classical
black hole singularities \RST . Consider a matter shock wave at ${x_0^+}$
as given by equation \fofour. Beneath the shock wave $({x^+ < x_0^+})$,
the geometry is the vacuum. The equations imply that ${\rho}$ and ${\phi}$, but
not their first derivatives ${\partial_+}{\rho}$ and ${\partial_+}{\phi}$,
are continuous across the shock wave. The geometry above the shock wave can
then be perturbatively computed
in a Taylor expansion about the
shock wave.  One finds that just above the shock wave \refs{\RST,\BDDO}
\eqn\phsk{{\partial_+}{\phi}(x_0^+,x^-) =
{{1}\over{{2}{x_0^+}}}\,\,({{M/\lambda}\over
{\sqrt{P}({\phi}({x_0^+},{x^-}))}} - {1}),}
\noindent where by continuity ${\phi}({x_0^+}, {x^-})$ is given by its
vacuum value.

There are two notable features of this expression. The first is that
${\partial_+}{\phi}$ diverges when the shock wave crosses the timelike line
in the vacuum where ${\phi} = {\phi_{cr}}$ (and ${P}$ vanishes). Before
diverging, however, it must cross zero at an earlier value ${x_H^-}$ of
${x^-}$. As discussed in section five, this point marks the beginning of
an apparent horizon. Behind this horizon and above the shock wave there
is a region of trapped points, or an apparent black hole. The
singularity at ${\phi} = {\phi_{cr}}$ is thus inside an apparent black hole.

In a region of trapped points lines of constant ${\phi}$ are spacelike.
Therefore the singularity at ${\phi} = {\phi_{cr}}$ leaves the shock
wave on a spacelike trajectory. It can also be seen \RST\ that the apparent
horizon leaves the shock wave on a timelike trajectory, corresponding
to the fact that the black hole is radiating and shrinking.

\ifig\fsus{Large-N black hole formation/evaporation. One possible
outcome is that the singularity and apparent horizon asymptotically
approach the event horizon at ${i^+}$ .}
{\hskip .5in \epsfysize=4.5in \epsfbox{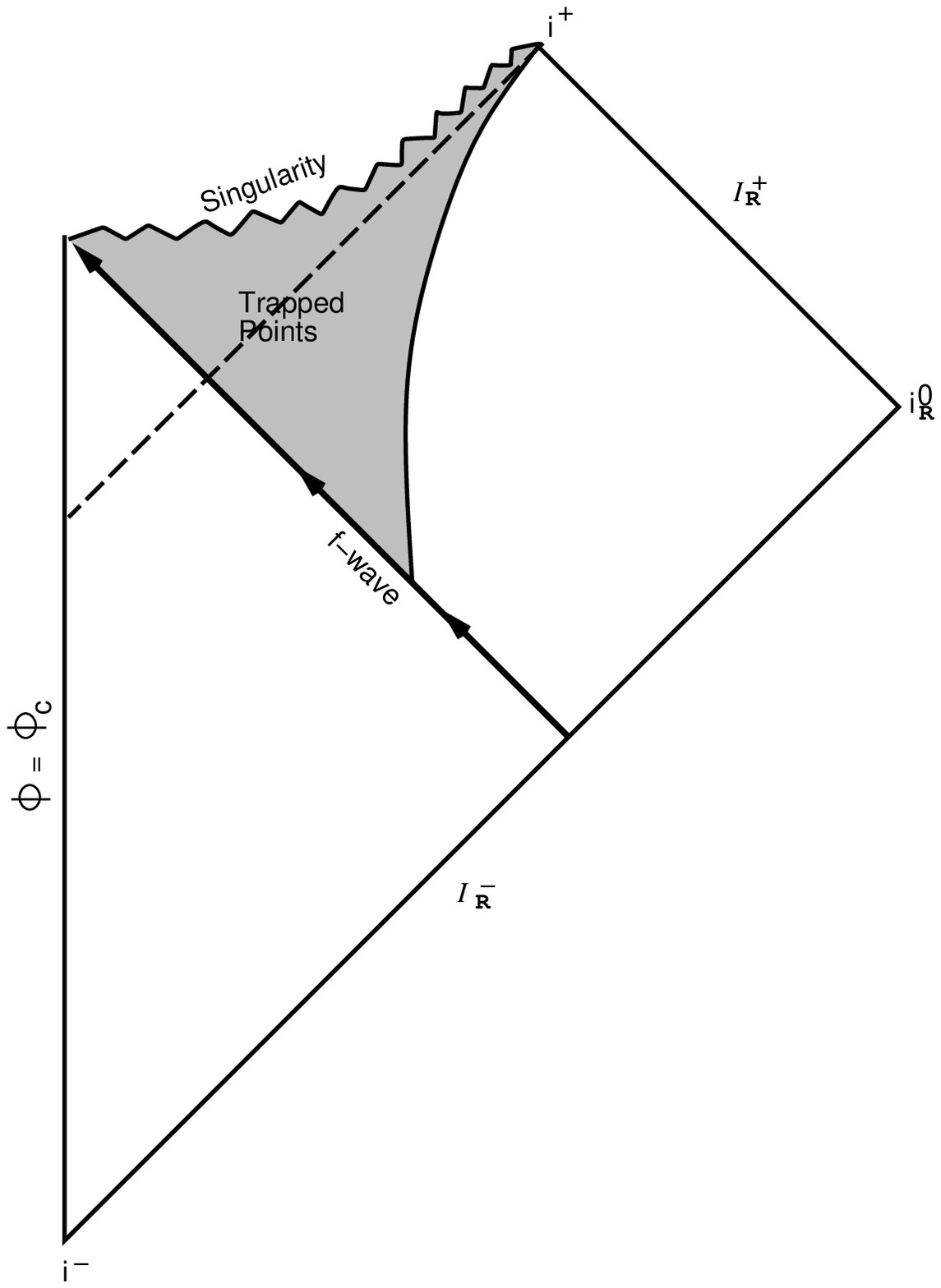}}

Unfortunately it is difficult to ascertain from the equations
the final fate of the apparent
black hole. It was initially believed \SuTh\ that the singularity and apparent
horizon were both asymptotic to a null line (the event horizon) which
meets ${\cal I}_R^+$ at future timelike infinity. This
is similar to the ``remnant'' alternative (II), and is depicted in
\fsus.

\ifig\fthn{Large-N black hole formation/evaporation. A second possible
outcome is a ``thunderbolt'' extending to ${\cal I}_R^+$. ${\cal I}_R^+$
is geodesically incomplete and ${i^+}$ never appears.}
{\hskip .5in \epsfysize=4.5in \epsfbox{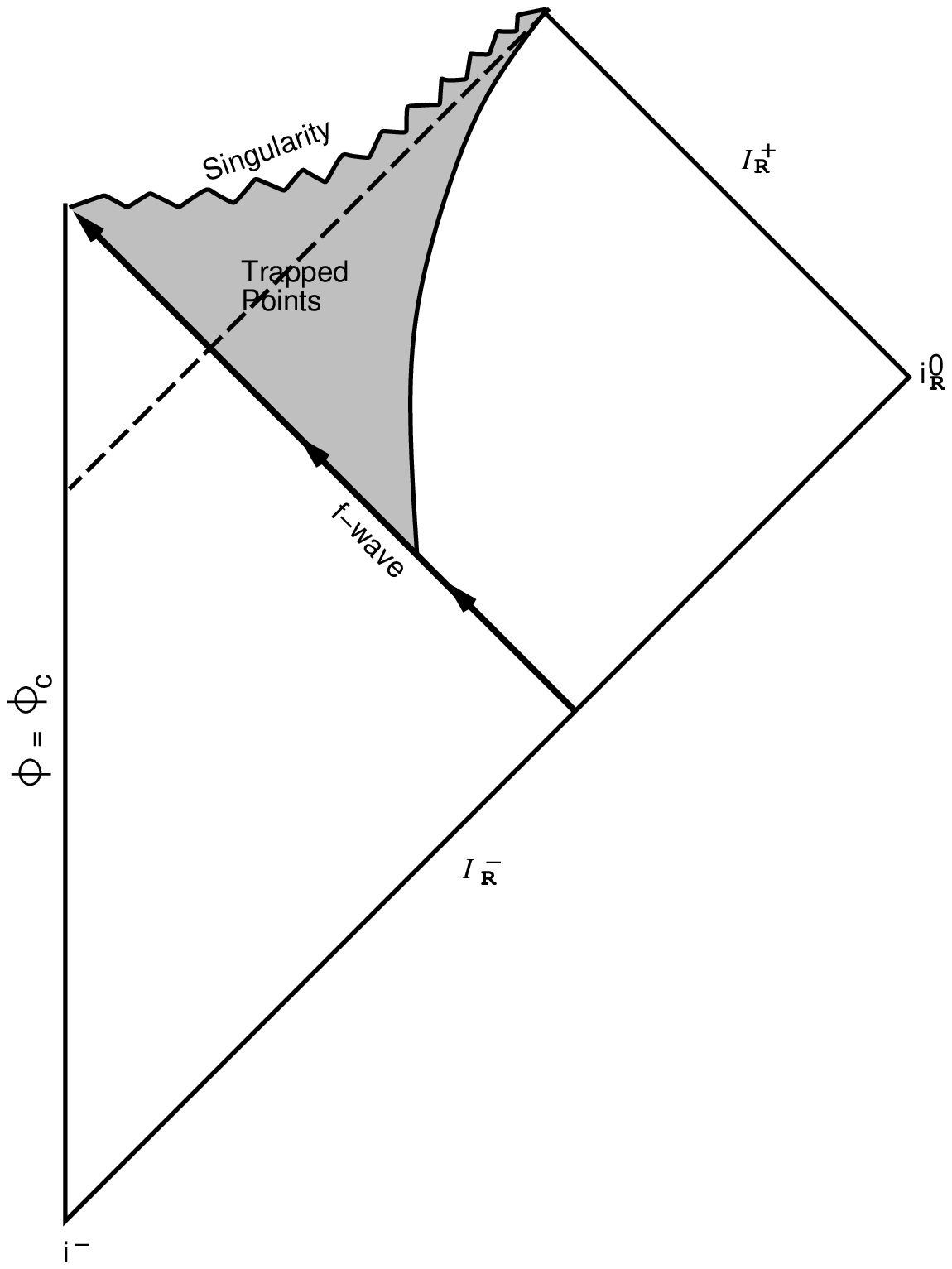}}

However \hwhat\  numerical evidence has been found
that the
event horizon meets ${\cal I}_R^+$ at a finite retarded time \hwhat. This is
depicted
in \fthn\ and corresponds to alternative (IV):  ``None of the above.''
While the causal structure looks similar, this has the crucial difference
that all timelike observers will run into the singularity -- which
Hawking refers to as a ``thunderbolt.''

\ifig\fdis{Large-N black hole formation/evaporation. A third possible
outcome is that the black disappears in a finite time, and the system reverts
to the vacuum.}
{\hskip .5in \epsfysize=4.5in \epsfbox{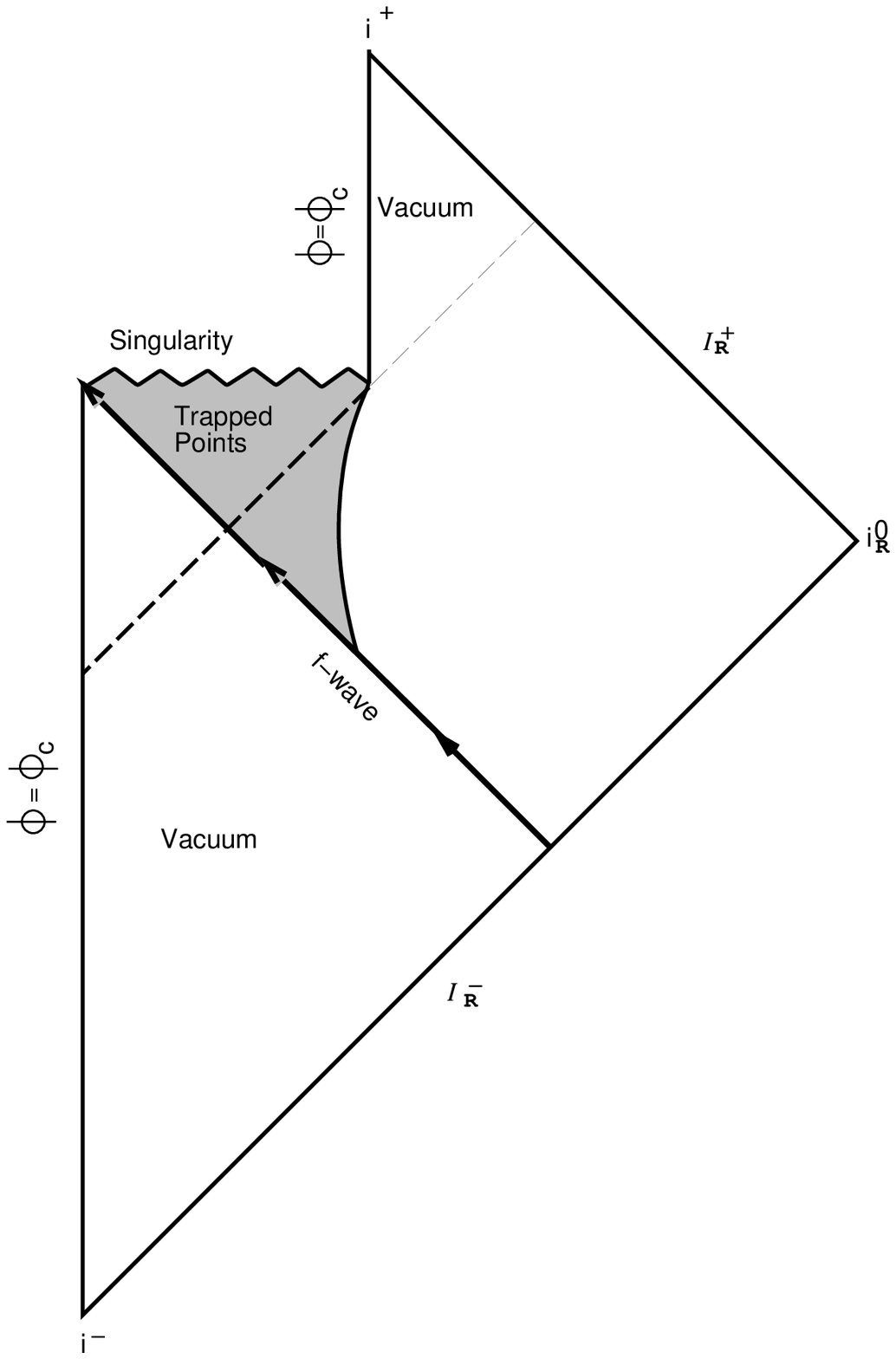}}

Yet a third possibility \andywaffle, illustrated in \fdis, is that the
apparent horizon
and the
singularity meet in a finite time, and a naked singularity appears.
Boundary conditions are then necessary to determine further evolution of the
system. Potentially these boundary conditions can be chosen to restore
the system to the vacuum\foot{This has indeed been found to be possible
at least for a modified version of the large-${N}$ equations \rst. Note that
other
boundary conditions could lead to a thunderbolt. }.
This would correspond to alternative (I).

It is important to stress that the large${-N}$ approximation can not
be trusted in regions where the fields themselves grow to be of order
${N}$. In particular the semiclassical equations must break down before
the singularity is reached, and one cannot reliably conclude that a singularity
does indeed exist. To probe the region near the ``singularity'' apparently
requires a non-perturbative treatment of the quantum theory.

Nevertheless, it should be possible to distinguish between the three
alternatives depicted in figs.12,13,14,
as they differ in regions where
${\phi}<<{\phi_{cr}}$. This is an important problem for future
research.

\subsec{${N} < {24}$ Gravitational Collapse}

In this subsection we will consider the one-loop semiclassical equations for
dilaton gravity with ${N} < {24}$ matter fields \Stro . The use of these
equations is not justified by a small parameter everywhere in the spacetime.
This is nevertheless a useful exercise because -- as will be seen -- it
illustrates that the singular behavior of the large ${N}$ equations may not
be universal, and indicates what other types of behavior may occur.

The one loop equations are of the form \phin\ and \rhon, except now with
\eqn\plp{{P} = {e^{-4\phi}} - {{N}\over{12}} {e^{-2\phi}} + {{N}\over{24}},}
\eqn\pplp{{P^{\prime}} = {-} {4} {e^{-4\phi}} + {{N}\over{6}} {e^{-2\phi}}.}
\noindent  The zeros of ${P}$ occur at
\eqn\ooo{{e^{-2\phi}} = {N \over 24}({1} {\pm} {\sqrt{{1} - {{24}\over{N}}})}.}
\noindent For ${N} > {24}$ there are two zeros which coalesce and move
into the complex plane for ${N} < {24}$. When ${N} < {24}$, ${P}$ is
an everywhere nonvanishing positive function.

\ifig\ffpbh{An $f$ shock wave incident on the vacuum. For $N<24$,
no singularities are encountered in a Taylor expansion about the shock
wave, but this does not probe the region above the dashed line .}
{\hskip .5in \epsfysize=3.5in \epsfbox{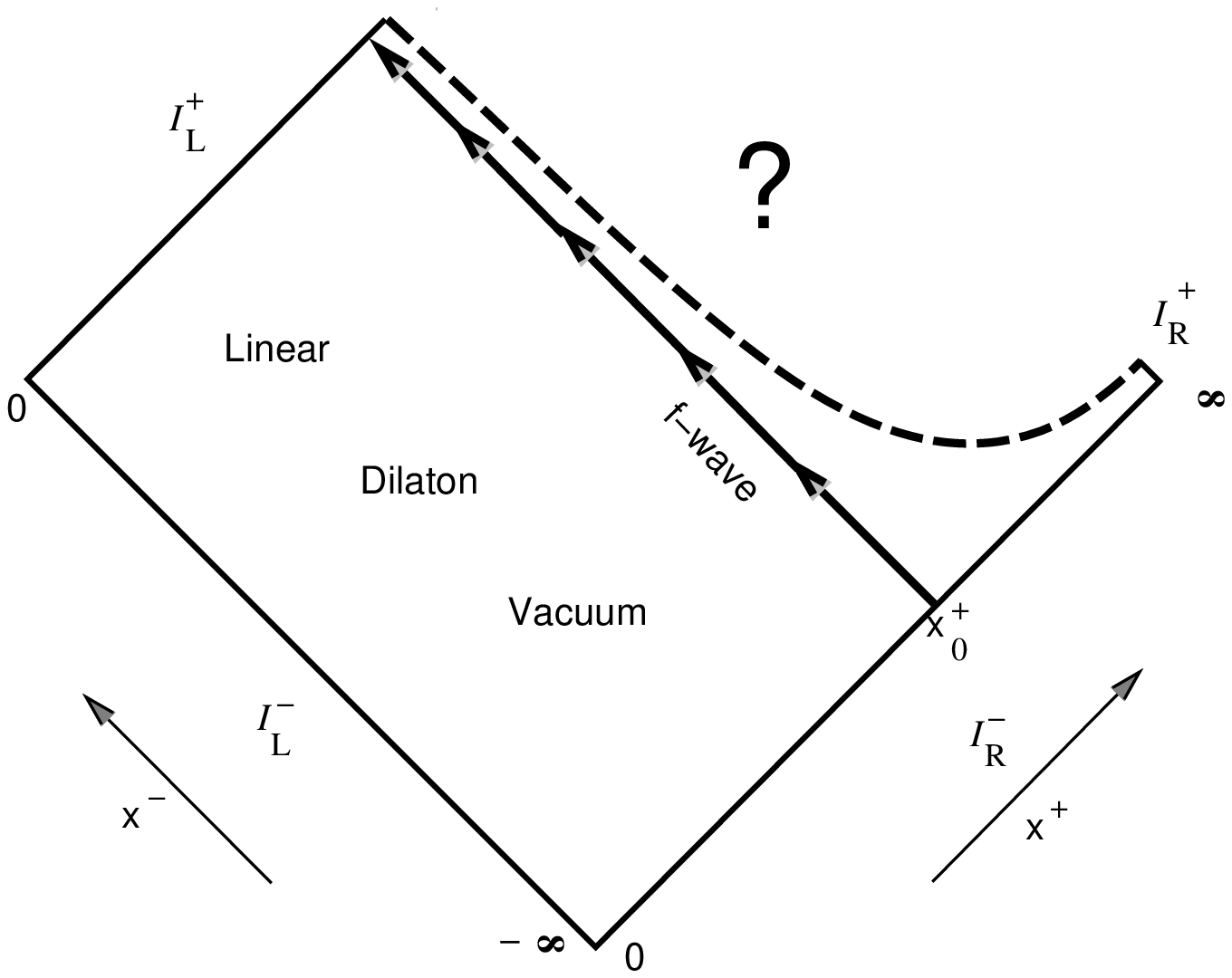}}

Thus there is no reason to expect singularities, and indeed none are found
in a Taylor expansion about the shock wave $({\partial_+}{\phi}$ is
given by \phsk\ with the modified ${P}$.) (The existence of singularities
outside the reach of the expansion about the shock wave has not been
ruled out.) Instead of producing a singularity, the shock wave surrounds
itself with a cloud of negative energy Hawking radiation, and arrives at
${\cal I}_L^+$ in a zero-energy bound state. This has the interesting
consequence that information can be deposited in ${\cal I}_L$ at zero cost
in energy. The situation is depicted in \ffpbh.

If there are indeed no singularities, then a unitary two-dimensional
${S}$-matrix from ${\cal I}^-_R \oplus {\cal I}^-_L$
to  ${\cal I}^+_R \oplus {\cal I}^+_L$
could exist. The four-dimensional interpretation of this
would be alternative (II), with the infinite degeneracy of zero-energy
bound states on ${{\cal I}_L}$ corresponding to the infinite remnant
species (referred to as cornucopions in \BDDO ).

\subsec{Quantum Black Holes}

Further evidence for the absence of singularities for ${N}<{24}$ can be found
by looking at the static one-loop equations.
Physically one does not expect the one-loop equations to contain
static, finite-mass black hole solutions. This is because these equations
include the effects of Hawking radiation, which destabilizes black
holes. However it is a curious feature of two-dimensional dilaton
gravity that one does have static solutions corresponding to ``quantum
black holes'' in thermal equilibrium with a radiation bath. This
radiation bath has a finite, constant energy density $N \lambda^2 /12$
extending all the way out to infinity\foot{Recall that the Hawking temperature
and
radiation energy density are independent of $M$ in two dimensions.}.
However this does not prevent
the black hole from being asymptotic to the vacuum, because the
gravitational coupling decreases exponentially at infinity. While the
deviation of a classical black hole geometry from the vacuum
decreases as ${Me^{-\lambda\sigma}}$ for large ${\sigma}$, that of a
quantum black hole goes as $(N \lambda^2 /12){\sigma}{e^{-\lambda\sigma}}$.
In four dimensions the radiation does not asymptotically decouple,
and there is no analogous solution.

Such static
quantum black hole solutions for large ${N}$ -- which is qualitatively
similar to one-loop ${N}>{24}$ -- were investigated numerically in
\refs{\Hawk,\BGHS}\ . The causal structure
is identical to that of a classical black hole and is
depicted in \ffive.
The main difference is that
the singularity now occurs at a finite value of ${\phi}$.

\ifig\fdesit{Penrose diagram for an $N<24$ quantum black hole
in equilibrium with a radiation bath.
The singularity seen in \ffive\ is replaced by an asymptotically deSitter
region.}{\hskip .5in\epsfysize=2.5in \epsfbox{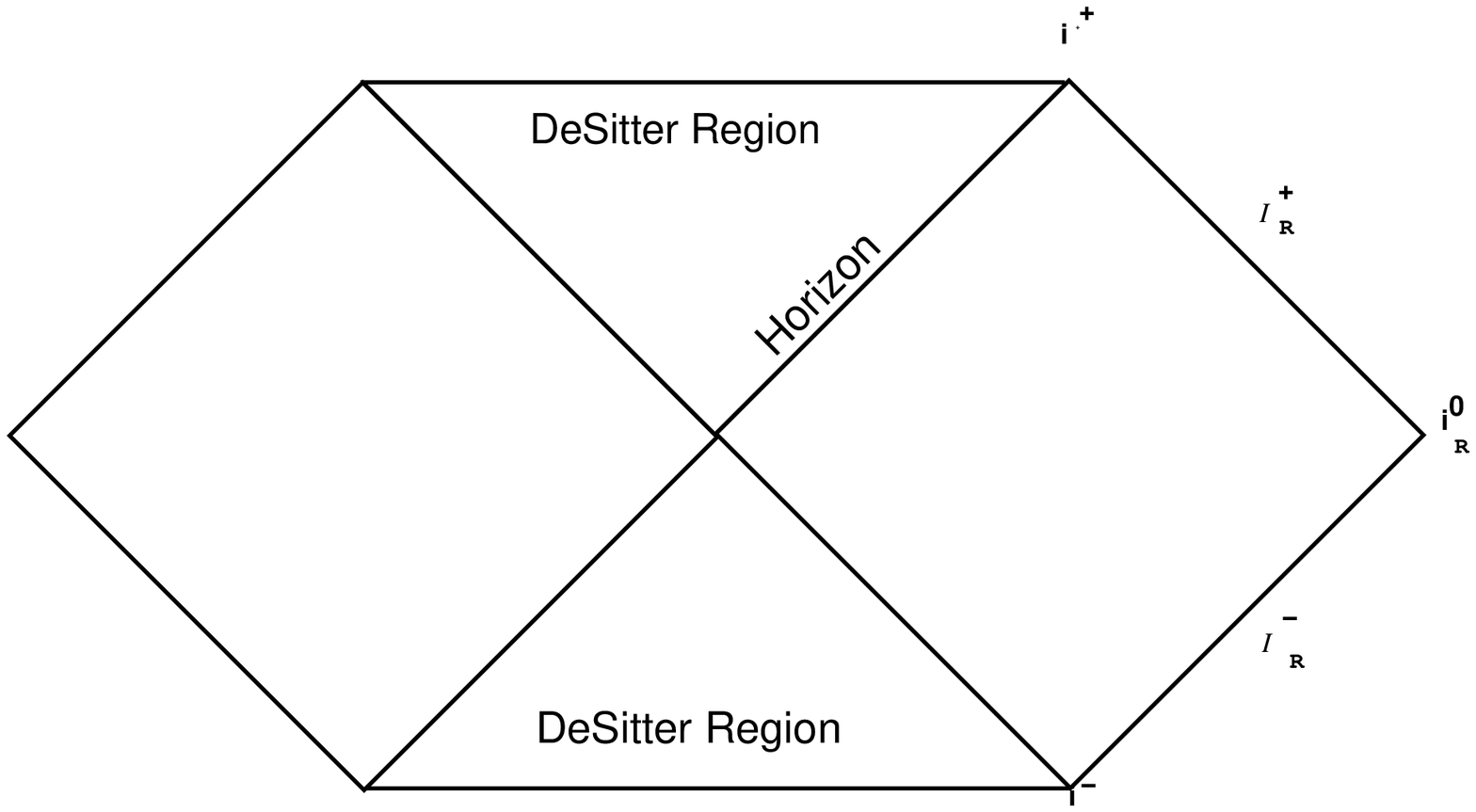}}

The ${N}<{24}$ equations exhibit markedly different behavior as illustrated
in \fdesit. There is
no singularity in the interior of the black hole. Instead it is
matched on to an asymptotically DeSitter region, with the future
(past) black hole horizon matched to the future (past) horizon of a
DeSitter observer at past (future) timelike infinity.

Clearly the ${N}<{24}$ equations are resistant to singularity formation.
Even if a black hole is forced into existence by pumping in energy as fast
as it is radiated out, no singularity is formed in the interior.

\newsec{Beyond One Loop}

While much was learned in the semiclassical analysis of the previous
sections, it was also clear that the answers to some questions must
await a more complete analysis of the quantum theory. Recent attempts
in this direction can be found in \refs{\tbo,\rut,\BiCa,\deAli,
\qtdg,\ham,\mik}. In  this section we will discuss some of the issues which
arise in
this regard.

The first point is that there is not a unique quantization of dilaton
gravity. If the quantum theory is defined as an expansion in ${e^{2\phi}}$,
there are new finite, renormalizable, counterterms at every order
in perturbation theory. For example at ${n}$th order there is the
term ${e^{{2}({n}-{1}){\phi}}}({\nabla\phi})^2$. While some constraints
on these terms will be discussed, they are far from being completely
fixed.

One elementary constraint is that the theory should have a stable ground
state. In fact it is quite easy to destabilize the ground state in the
process of adding
terms to the action. General criteria for the existence of a positive
energy theorem are discussed in \past .

Further properties of the quantum theory follow from the connection between
two-dimensional gravity and conformal field theory \jptb. This connection
is best understood by quantizing the theory in conformal gauge:
\eqn\eee{\eqalign{{g_{+-}}&= {-} \half\,{e^{2\rho}},\cr
{g_{++}}&= {g_{--}} = {0}.\cr}}
\noindent This gauge leaves unfixed a group of residual diffeomorphisms
for which
\eqn\hhh{\eqalign{
{\delta}{g_{++}}&= {\nabla_+}{\zeta_+} = {g_{+-}}{\partial_+}
{\zeta^-} = {0},\cr
{\partial}{g_{--}}&= {\nabla_-}{\zeta_-} = {g_{+-}}{\partial_-}{\zeta^+}
= {0}.\cr}}
\noindent These equations imply
\eqn\YYY{{\zeta^{\pm}} = {\zeta^{\pm}} ({\sigma^{\pm}}),}
\noindent and that the residual diffeomorphisms generate the conformal
group. Correspondingly the moments of ${T_{++}}$ and ${T_{--}}$ generate
Virasoro algebras.

Invariance of the quantum theory under the residual
symmetry group can be insured, for example, by constructing a BRST charge
${Q}$ which obeys ${Q^2} = {0}$ and identifying physical states as
${Q} $ - cohomology classes.

At this point it should be clear that -- although a slightly different set of
words is being used -- what is being constructed here is a ${c} = {26}$
conformally invariant sigma model with ${\rho}, {\phi}$ and ${f_i}$
as fields living in an ${N} + {2}$ dimensional target space. If one
demands that the matter fields ${f_i}$ constitute a free ${c} = {N}$ conformal
field theory, then the ${\rho}, {\phi}$ sigma model must be conformally
invariant with ${c} = {26} - {N}$.

Letting ${X^\mu} = ({\rho}, {\phi})$, the ${\rho}, {\phi}$ sigma model
can be written in the form:
\eqn\FFF{
{S} = {-} {{{1}\over{2\pi}}} {\int} {d^2}{x} {\sqrt{{-}{\hat g}}} [{\cal G}
_{\mu\nu}{\nabla}{X^\mu}{\nabla}{X^\nu}
+ {{1}\over{2}}
{\Phi}{\hat R} + {T}],}
\noindent ${\hat g}$ here is a fiducial metric and ${\cal G}$, ${\Phi}$
and ${T}$ are functions of ${X^\mu}$. The couplings ${\cal G},
{\Phi}$ and ${T}$ are severely restricted by conformal
invariance. Namely, the beta functions must vanish:
\eqn\bfuns{\eqalign{0&=\beta^G_{\mu\nu}  = {\cal R}_{\mu\nu}+
2\nabla_\mu\nabla_\nu \Phi+\cdots , \cr
                      0&=     \beta^\Phi ={N-24 \over 3}  - \frac{{\cal R}}{4}
+
              {(\nabla\Phi)^2} - \nabla^2\Phi+\cdots ,\cr
                            0&= \beta^T   = \nabla^2 T
-2\nabla \Phi\cdot \nabla T +8T+\cdots\cr}}
where ${\cal R}$ is the curvature of $\cal G$.
These equations are indeed obeyed, to leading order, by the
${\cal G}, {\Phi}$ and ${T}$ implicit in section eight.
While conformal invariance severely constrains the quantum theory, there are
still an infinite number of solutions. This may be viewed as an initial
data problem in which initial data is specified as a function of
${\phi}$ at fixed ${\rho}$, and the beta function equations are
then used to solve for ${\cal G}, {\Phi}$ and ${T}$ at every
value of ${\rho}$.

In order to correspond to the theory of dilaton gravity that we are
interested in, the values of ${\cal G},~{\Phi}$ and ${T}$ at
weak coupling $({\phi}{\rightarrow}{-}{\infty})$ should agree with
those appearing in our one-loop expressions in section eight.

It is natural to ask whether ${\cal G},~{\Phi}$ and ${T}$ can
be specified in such a way so as to correspond to an exactly
solvable conformal field theory. In \refs{\BiCa,\deAli,\rut}\
it was shown that if ${\cal G}$ and ${\Phi}$ are given exactly by their
one-loop expressions, ${T}$ can be adjusted at higher orders
in such a way as to obtain a soluble Liouville-like conformal field
theory. This construction manifestly gives a theory which agrees
with dilaton gravity at weak coupling. Unfortunately, the required
adjustments of ${T}$ destabilize the vacuum \qtdg . While this preliminary
effort did not quite succeed, perhaps some modification will.

\newsec{Conclusion}

The phenomena of black hole evaporation raises a deep question regarding
the compatibility of quantum mechanics and gravity. Do quantum mechanical
wave functions evolve into
a definite future, or can they only be predicted statistically? In this
review we have explained how this same question arises in two-dimensional
models, which are computationally much more tractable than their
four-dimensional partners. There is a large class of relevant
two-dimensional models, and only a few have been investigated so far.
Further investigations of these models, together with their connection
to our four-dimensional world, holds the promise of revealing new insights
about the nature of our own future.

\centerline{\bf Acknowledgments}

We are grateful to many people for conversations on black holes
including Tom Banks, Bjorn Birnir, David Garfinkle, Ruth Gregory,
Stephen Hawking,
Gary Horowitz, Emil Martinec, Greg Moore, Don Page,  Leonard Susskind, Larus
Thorlacius, Herman Verlinde and Bob Wald. We are particularly grateful to
Curt Callan and
Steve Giddings for collaboration and numerous
discussions, and to John Preskill
for arousing our interest in the problem of particle-hole scattering.
We would also like to thank the ICTP Trieste and the TASI
local organizers for their hospitality.
This work was supported in part by DOE Grant No. 91ER40618 and NSF Grant
No. PHY90-00386. J.H. also acknowledges the support of NSF PYI
Grant No. PHY-9196117.
\listrefs
\end